\pdfoutput=1

\documentclass[twocolumn, twocolappendix]{aastex63}

\newcommand{\beq}{\begin{equation}}
\newcommand{\beqa}{\begin{eqnarray}}
\newcommand{\eeq}{\end{equation}}
\newcommand{\eeqa}{\end{eqnarray}}

\newcommand{\simgt}{\lower.5ex\hbox{$\; \buildrel > \over \sim \;$}}
\newcommand{\simlt}{\lower.5ex\hbox{$\; \buildrel < \over \sim \;$}}

\newcommand{\bd}[1]{\mbox{\boldmath $#1$}}

\newcommand{\ms}[1]{\textcolor{black}{#1}}

\submitjournal{ApJ}

\shorttitle{Halo mass function in Planck cosmology}
\shortauthors{Shirasaki et al.}
\graphicspath{{./}{figures/}}

\begin{document}

\title{Virial halo mass function in the {\it Planck} cosmology}

\correspondingauthor{Masato Shirasaki}
\email{masato.shirasaki@nao.ac.jp}


\author{Masato Shirasaki}
\affiliation{National Astronomical Observatory of Japan, Mitaka, Tokyo 181-8588, Japan}
\affiliation{The Institute of Statistical Mathematics, Tachikawa, Tokyo 190-8562, Japan}


\author{Tomoaki Ishiyama}
\affiliation{Institute of Management and Information Technologies, Chiba University, Chiba, 263-8522, Japan}

\author{Shin'ichiro Ando}
\affiliation{GRAPPA Institute, University of Amsterdam, 1098 XH Amsterdam, The Netherlands}
\affiliation{Kavli Institute for the Physics and Mathematics of the Universe (WPI), University of Tokyo, Chiba 277-8583, Japan}



\begin{abstract}
We study halo mass functions with high-resolution $N$-body simulations under a $\Lambda$CDM cosmology.
Our simulations adopt the cosmological model that is consistent with recent measurements of the cosmic microwave backgrounds with the {\it Planck} satellite.
We calibrate the halo mass functions for $10^{8.5} \lower.5ex\hbox{$\; \buildrel < \over \sim \;$} M_\mathrm{vir} / (h^{-1}M_\odot) \lower.5ex\hbox{$\; \buildrel < \over \sim \;$} 10^{15.0 - 0.45 \, z}$,
where $M_\mathrm{vir}$ is the virial spherical overdensity mass 
and redshift $z$ ranges from $0$ to $7$.
The halo mass function in our simulations can be fitted by a four-parameter model 
over a wide range of halo masses and redshifts, while we require some redshift evolution of 
the fitting parameters. Our new fitting formula of the mass function has a 5\%-level precision except for 
the highest masses at $z\le 7$. Our model predicts that the analytic prediction in Sheth \& Tormen would overestimate the halo abundance at $z=6$ with $M_\mathrm{vir} = 10^{8.5-10}\, h^{-1}M_\odot$ by 20--30\%. Our calibrated halo mass function provides a baseline model to constrain warm dark matter (WDM) by high-$z$ galaxy number counts. 
We compare a cumulative luminosity function of galaxies at $z=6$
with the \ms{total} halo abundance based on our model and a recently proposed WDM correction.
We find that WDM with its mass lighter than $2.71\, \mathrm{keV}$ is incompatible with the observed galaxy number density at a $2\sigma$ confidence level.
\end{abstract}



\section{Introduction} \label{sec:intro}

Understanding the origin and evolution of large-scale structures 
is one of the most important subjects in modern cosmology.
In the current standard cosmological model, 
referred to as the $\Lambda$ cold dark matter ($\Lambda$CDM) model,
the formation of astronomical objects is expected to occur hierarchically.
Dark matter halos are gravitationally bound objects 
made through non-linear evolution of cosmic mass density.
Halos can compose the large-scale structures in the Universe.
Since galaxies would be born in dark matter halos \citep[e.g.][]{1978MNRAS.183..341W, 1999MNRAS.310.1087S, 2015ARA&A..53...51S},
the abundance of halos plays a central role 
in understanding statistical properties of observed galaxies in modern large surveys.

Mass function of dark matter halos is defined by the halo abundance 
as a function of halo masses.
There are various application examples of the halo mass function in practice.
Those include constraining cosmological parameters 
with a number count of galaxy clusters \citep[e.g.][for a review]{2011ARA&A..49..409A}
and inference of the relation between stellar and total masses in single galaxies
\citep[e.g.][for a review]{2018ARA&A..56..435W}.

\begin{figure*}[!t]
\gridline{\fig{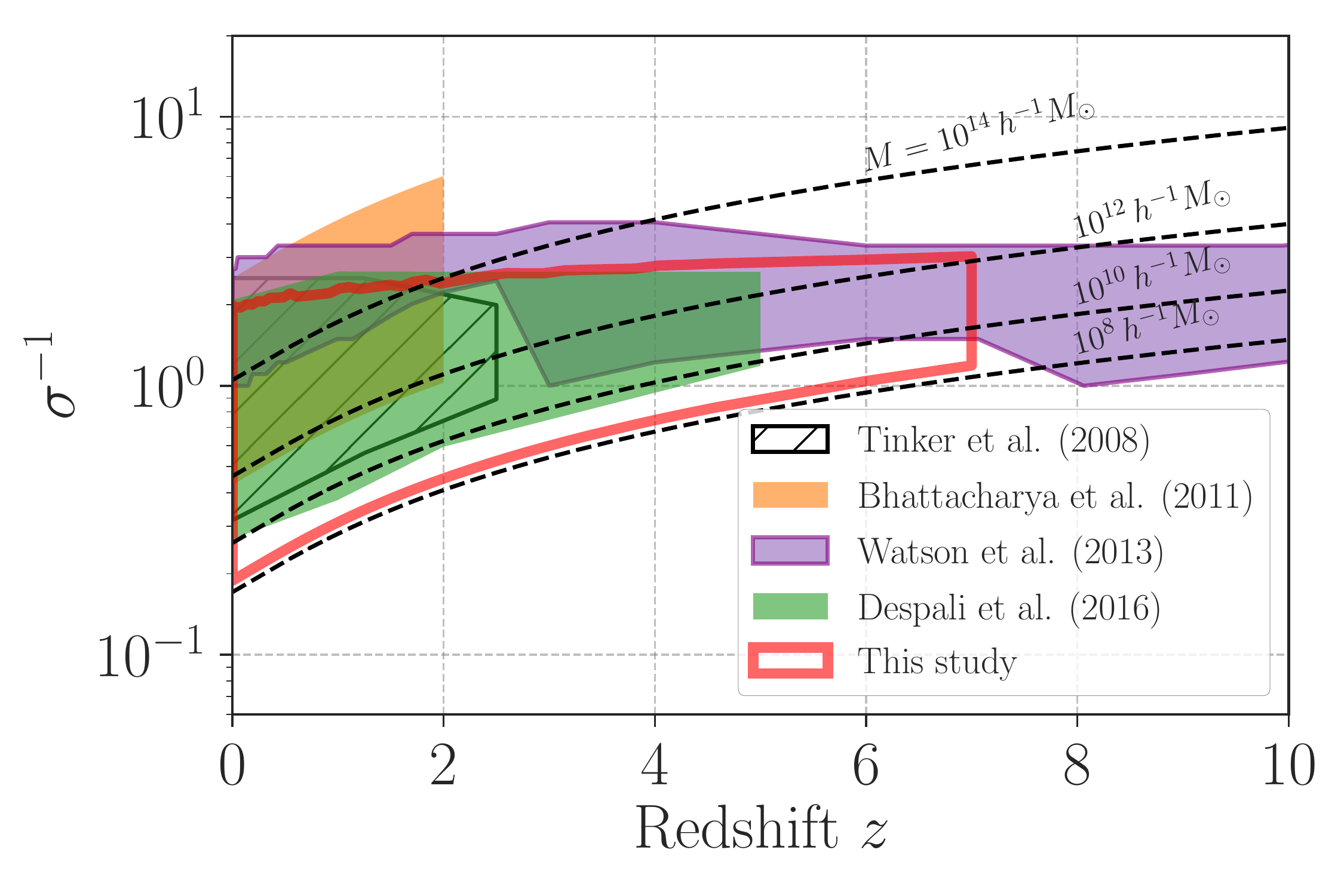}{0.80\textwidth}{}}
\caption{
Top-hat mass variances $\sigma^2$ covered by several $N$-body simulations.
The red \ms{open} region represents the parameter space explored in this paper.
Our paper aims at calibrating the halo mass function at lower-mass regimes ($M\simgt 10^{8}\, h^{-1}M_{\odot}$) than previous studies at redshifts of $z\simlt7$.
For comparison, the shaded, orange filled, purple filled, and green filled regions show
the coverage in the $\sigma^{-1}$-$z$ plane in \citet{2008ApJ...688..709T},
\citet{2011ApJ...732..122B}, \citet{2013MNRAS.433.1230W}, and \citet{2016MNRAS.456.2486D}, respectively.
Note that black dashed lines provide the linear-theory predictions 
at $M=10^{14}\, h^{-1}M_{\odot}$, 
$10^{12}\, h^{-1}M_{\odot}$, $10^{10}\, h^{-1}M_{\odot}$ and $10^{8}\, h^{-1}M_{\odot}$ 
from top to bottom.
\label{fig:sigma_M_space}}
\end{figure*}

Although the formation of dark matter halos is governed 
by complex gravitational processes, there exist simple analytic predictions of the halo mass function \citep[e.g.][]{1974ApJ...187..425P, 1991ApJ...379..440B, 2002MNRAS.329...61S}.
The basic assumption in analytic approaches is that 
one can relate halos with their mass of $M$ with 
the linear density field smoothed at some scales of $R$.
A common choice of the scale $R$ is set to $R=(3M/4\pi\bar{\rho}_\mathrm{m})^3$
where $\bar{\rho}_\mathrm{m}$ is the average cosmic mass density.
The variance of the linear density field smoothed by a top-hat filter with $R$, 
denoted as $\sigma^2(R)$, is then used to characterize the mass fraction of dark matter halos of $M$.
Under simple but physically-motivated assumptions,
the analytic approaches predict that any dependence of halo masses, redshifts, 
and underlying cosmological models in the halo mass function 
can be determined by the variance $\sigma(R)$ alone.
One can factor out a pure $\sigma$ dependence in the halo mass function 
with the analytic approaches.
This $\sigma$ dependence is known as the multiplicity function $f(\sigma)$,
expecting that it is a universal function for different halo masses, redshifts, and cosmological models.

Numerical simulations have validated the universality of the multiplicity function so far.
\citet{2001MNRAS.321..372J} constrained the redshift- and cosmology-dependence of the multiplicity function to be less than a $\sim15\%$ level when the halo mass is defined by the Friends-of-friends (FoF) algorithm \citep{1985ApJ...292..371D} with some FoF linking lengths.
Different definitions of halo masses can introduce a systematic non-universality of the multiplicity functions in simulations \citep{2002ApJS..143..241W, 2008ApJ...688..709T, 2020ApJ...903...87D}.
As increasing particle resolutions, several groups have found that the multiplicity function for the FoF halos depends on cosmological models with a 10\% level \citep[e.g.][]{2006ApJ...646..881W, 2011ApJ...732..122B} but its redshift dependence is weak \citep[e.g.][]{2013MNRAS.433.1230W}.
Recently, \citet{2016MNRAS.456.2486D} have claimed the universality of the multiplicity function in their simulation sets when defining the halo mass with a virial spherical overdensity
and expressing the multiplicity function in terms of a re-scaled $\sigma$.

In this paper, we extend previous measurements of the multiplicity function 
toward lower halo masses.
Figure~\ref{fig:sigma_M_space} summarizes 
the coverage of halo masses and redshifts in our paper. 
In the figure, we convert the halo mass scale $M$ to 
the top-hat variance $\sigma^2(R)$ using the linear
matter power spectrum in the best-fit $\Lambda$CDM cosmology inferred in \citet[][Planck16]{2016A&A...594A..13P}.
The red shaded region in Figure~\ref{fig:sigma_M_space} shows our coverage, while
other shaded and hatched regions represent ones in some of previous studies.
Our measurements of $f(\sigma)$ include the range of 
$10^{8.5} \le M / (h^{-1}M_\odot)\simlt 10^{10}$ at wide redshifts of $z\le 7$, 
which is not explored in the literature.
We examine the universality of the multiplicity function in the Planck16 $\Lambda$CDM cosmology
from gaseous mini-halos \citep[e.g.][]{2017MNRAS.465.3913B, 2020MNRAS.498.4887B} to massive galaxy clusters.
To do so, we analyze dark-matter-only $N$-body simulations with different resolutions
at 40 different redshifts in \citet{2015PASJ...67...61I, 2020MNRAS.492.3662I}, allowing to study possible redshift evolution of the multiplicity function in details.

The paper is organized as follows. 
In Section~\ref{sec:data}, we describe the simulation data of 
dark matter halos used in this paper.
In Section~\ref{sec:model},
we present an overview of the halo mass function,
introduce how to estimate the multiplicity function from simulated halos
as well as statistic errors in our measurements.
The analysis pipeline to find the best-fit model to our measurements is provided in Section~\ref{sec:calibration}.
The results are presented in Section~\ref{sec:results}, while we summarize some limitations of our results in Section~\ref{sec:limitations}.
Finally, the conclusions and discussions are provided in Section~\ref{sec:conclusion}.
Throughout this paper, we assume the cosmological parameters in Planck16.
To be specific, we adopt
the cosmic mass density $\Omega_{\mathrm{m0}}= 0.31$,
the baryon density $\Omega_{\mathrm{b0}}= 0.048$,
the cosmological constant $\Omega_{\Lambda}=1-\Omega_{\mathrm{m0}} = 0.69$,
the present-day Hubble parameter $H_0 = 100h\, \mathrm{km}\,\mathrm{s^{-1}}\,\mathrm{Mpc^{-1}}$ with $h= 0.68$,
the spectral index of primordial curvature perturbations 
$n_s= 0.96$, and the linear mass variance smoothed over $8\, h^{-1}\mathrm{Mpc}$, $\sigma_8= 0.83$.
We also refer to $\log$ as the logarithm with base 10, while $\ln$ represents the natural logarithm.

\section{$N$-body simulations and halo catalogs} \label{sec:data}

To study the abundance of dark matter halos, 
we use a set of halo catalogs based on high-resolution cosmological 
$N$-body simulations with various combinations of mass resolutions and volumes \citep{2015PASJ...67...61I, 2020MNRAS.492.3662I}.
In this paper, we use the halo catalogs based on three different runs of
$\nu^2$GC-L (L), $\nu^2$GC-H2 (H2), and phi1.\footnote{The halo catalogs are available at \url{https://hpc.imit.chiba-u.jp/~ishiymtm/db.html}.
\ms{Note that $\nu^2$GC stands for \underline{new} \underline{nu}merical \underline{g}alaxy \underline{c}atalogs.}}
Table~\ref{tab:table_sims} summarizes specifications of our simulation sets.

The simulations were performed by a massive parallel TreePM code of 
{\tt GreeM}\footnote{\url{https://hpc.imit.chiba-u.jp/~ishiymtm/greem/}} 
\citep{2009PASJ...61.1319I, 2012arXiv1211.4406I} on the K computer at the RIKEN Advanced Institute for Computational Science, 
and Aterui super-computer at Center for Computational Astrophysics (CfCA) of 
National Astronomical Observatory of Japan.
The authors generated the initial conditions for the L and H2 runs 
by a publicly available code, {\tt 2LPTic},\footnote{\url{http://cosmo.nyu.edu/roman/2LPT/}}
while another public code {\tt MUSIC}\footnote{\url{https://bitbucket.org/ohahn/music/}} 
\citep{2011MNRAS.415.2101H} has been adopted to generate the initial conditions for 
the phi1 run.
Note that either public code uses second-order Lagrangian perturbation theory \citep[e.g.][]{2006MNRAS.373..369C}.
All simulations began at $z=127$.
The linear matter power spectrum at the initial redshift has been computed with
the online version of {\tt CAMB}\footnote{\url{http://lambda.gsfc.nasa.gov/toolbox/tbcambform.cfm}} \citep{Lewis:1999bs}.
In the simulations, the Planck16 cosmological model has been adopted.

\begin{table}[t!]
\renewcommand{\thetable}{\arabic{table}}
\centering
\caption{A summary of simulation sets analyzed in this paper. 
The total number of $N$-body particles ($N_p$),
the simulation box size on each side ($L_\mathrm{box}$),
the mass of each $N$-body particle ($m_p$),
and the softening length ($\epsilon$)
are provided for three different runs.
Note that the softening length is set to be $3\%$ of the mean free path of particles in each run.
} \label{tab:table_sims}
\begin{tabular}{c|c|c|c|c}
\tablewidth{0pt}
\hline
\hline
Name
&
$N_p$
&
$L_\mathrm{box}\, (h^{-1}\, \mathrm{Mpc})$
& 
$m_p\, (h^{-1}\, M_\odot)$
&
$\epsilon\, (h^{-1}\, \mathrm{kpc})$
\\
\hline
\decimals
L 
& $8192^3$
& $1120$
& $2.20\times10^{8}$
& $4.27$
\\
\hline
\decimals
H2 
& $2048^3$
& $70$
& $3.44\times10^{6}$
& $1.07$
\\
\hline
\decimals
phi1 
& $2048^3$
& $32$
& $3.28\times10^{5}$
& $0.48$
\\
\hline
\end{tabular}
\end{table}

All halo catalogs in this paper have been produced with 
the {\tt ROCKSTAR} halo finder\footnote{\url{https://bitbucket.org/gfcstanford/rockstar}}
\citep{2013ApJ...762..109B}.
We focus on parent halos identified by the {\tt ROCKSTAR} algorithm and 
exclude any subhalos in the following analyses. 
We keep the halos with their mass greater than 1000 times $m_p$,
where $m_p$ is the particle mass in the $N$-body simulations.
Throughout this paper, the halo mass is defined 
by a spherical virial overdensity \citep{1998ApJ...495...80B}.
We analyze the halo catalogs at 40 different redshifts below: 
$z=$0.00,
0.03, 0.07, 0.13, 0.19, 0.24, 0.30,
0.36, 0.42, 0.48, 0.55,
0.61,
0.69,
0.76,
0.84,
0.92,
1.01,
1.10,
1.20,
1.29,
1.39,
1.49,
1.60,
1.70,
1.83,
1.97,
2.12,
2.28,
2.44,
2.60,
2.77,
2.95,
3.14,
3.37,
3.80,
4.04,
4.29,
4.58,
5.98
and
7.00. 

It is worth noting that we adopt the default halo mass definition in the {\tt ROCKSTAR} finder.
This mass definition does not include unbound particles. 
Because unbound particles around a given dark matter halo can contribute to the spherical halo mass,
the halo mass function without unbound particles may contain additional systematic uncertainties.
In Appendix~\ref{apdx:strict_so_test}, we examine the impact of unbound particles in our measurements using a different set of $N$-body simulations. 
We confirmed that unbound particles did not introduce 
systematic errors in the calibration beyond statistic uncertainties.

\section{Halo Mass Function} \label{sec:model}

\subsection{Model}

In spite of a strong dependence on the matter power spectrum,
there exist successful analytical approaches predicting
the number density of dark matter halos \citep[e.g.][]{1974ApJ...187..425P, 1991ApJ...379..440B, 2002MNRAS.329...61S}.
Those approaches commonly relate the halos with their mass $M$ to the linear density field 
smoothed at some scale $R = (3M/4\pi\bar{\rho}_\mathrm{m})^{1/3}$.
To develop an analytic formula of the halo abundance, 
\citet{1974ApJ...187..425P} assumed
that the fraction of mass in halos of mass greater than $M$ at redshift $z$ 
is set to be twice the probability that smoothed Gaussian density fields exceed 
the critical threshold for spherical collapse, $\delta_{c}$.
In this ansatz, the number density of halos can be written as
\beqa
\frac{\mathrm{d}n}{\mathrm{d}M} = 
\frac{\bar{\rho}_\mathrm{m}}{M} \frac{\mathrm{d}\ln \sigma^{-1}}{\mathrm{d}M}\, f(\sigma), \label{eq:dndm}
\eeqa
where $\sigma$ is the root mean square (RMS) fluctuations of the linear density field smoothed with a filter encompassing this mass $M$. The RMS is usually defined with a spherical top-hat filter,
\beqa
\sigma^2(M, z) = \int \frac{k^2\mathrm{d}k}{2\pi^2} \, W^2_\mathrm{TH}(kR)\, P_\mathrm{L}(k,z), \label{eq:sigma_TH}
\eeqa
where $P_\mathrm{L}(k,z)$ is the linear matter power spectrum as a function of wavenumber $k$ and redshift $z$, and $W_\mathrm{TH}$ is the Fourier transform of the real-space top-hat window function of radius $R$.
To be specific, the top-hat window function is given by $W_\mathrm{TH}(x) = 3[\sin x-x \cos x]/x^3$.
\citet{1974ApJ...187..425P} found that the function $f$, referred to as the multiplicity function, 
is expressed as
\beqa
f_\mathrm{PS} = \sqrt{\frac{2}{\pi}}\frac{\delta_c}{\sigma}\exp\left(-\frac{\delta^2_c}{2\sigma^2}\right).
\eeqa
The multiplicity function in \citet{1974ApJ...187..425P} has been revised later 
by introducing excursion set theory \citep{1991ApJ...379..440B} and adopting the ellipsoidal collapse \citep{2002MNRAS.329...61S}.
Note that previous analytic models predict that the multiplicity function is a universal function
and any dependence on redshifts and cosmological models can be encapsulated in $\sigma(M,z)$.

Motivated by these analytic predictions, we assume that 
the multiplicity function can be described by a four-parameter model,
\beqa
f(\sigma, z) &=& A(z)\, \sqrt{\frac{2}{\pi}}\, \exp\left(-\frac{1}{2}\frac{a(z)\delta^2_{c,z}}{\sigma^2}\right)
\nonumber \\
&& 
\quad \times
\left[1+\left(\frac{\sqrt{a(z)}\delta_{c,z}}{\sigma}\right)^{-2p(z)}\right]\, \left(\frac{\sqrt{a(z)}\delta_{c,z}}{\sigma}\right)^{q(z)},
\label{eq:fmodel}
\eeqa
where $A(z)$, $a(z)$, $p(z)$, and $q(z)$ are free parameters in the model. In Eq.~(\ref{eq:fmodel}), we introduce the redshift-dependent critical overdensity for spherical collapse, $\delta_{c,z}$.
In a flat $\Lambda$CDM cosmology, this quantity is well approximated as \citep{1996ApJ...469..480K}
\beqa
\delta_{c,z} &=& \frac{3\, (12\pi)^{2/3}}{20}\, 
\left[1+0.123 \log\, \Omega_\mathrm{m}(z)\right], \\ \label{eq:delta_cz}
\Omega_\mathrm{m}(z) &=& \frac{\Omega_\mathrm{m0}(1+z)^3}{\Omega_\mathrm{m0}(1+z)^3+(1-\Omega_\mathrm{m0})}.
\eeqa

\subsection{Estimator of the multiplicity function}\label{subsec:estimator}

To find a best-fit model of $f(\sigma,z)$ to the simulation data, 
we need to construct the multiplicity function from the halo catalogs.
We start with a binned halo mass function, which is directly observable from the simulations.
Let $\Delta n_\mathrm{bin}$ be the comoving number density of dark matter halos in a bin size of 
$\Delta \log M$ with mass ranges $[M_1, M_2]$.
Using Eq.~(\ref{eq:dndm}), one finds 
\beqa
\Delta n_\mathrm{bin} = \ln 10\, \int_{\log M_1}^{\log M_2}\, \mathrm{d}\log M\,\frac{\bar{\rho}_\mathrm{m}}{M}\, 
\frac{\mathrm{d}\log \sigma^{-1}}{\mathrm{d}\log M}\, f(\sigma). \label{eq:nbin}
\eeqa
In the limit of $\Delta \log M \rightarrow 0$, we obtain
\beqa
f_\mathrm{sim}(\sigma)|_{M=M_\mathrm{bin}} 
&=& \frac{1}{\ln 10}\frac{M_\mathrm{bin}}{\bar{\rho}_\mathrm{m}}
\frac{\Delta n_\mathrm{bin}}{\Delta \log M} \nonumber \\
&&
\qquad \times \left(\frac{\mathrm{d}\log \sigma^{-1}}{\mathrm{d}\log M}\right)^{-1}\Biggr{|}_{M=M_\mathrm{bin}}, \label{eq:fsim}
\eeqa
where $M_\mathrm{bin}$ is a center of the binned mass.

\begin{figure}[t!]
\gridline{\fig{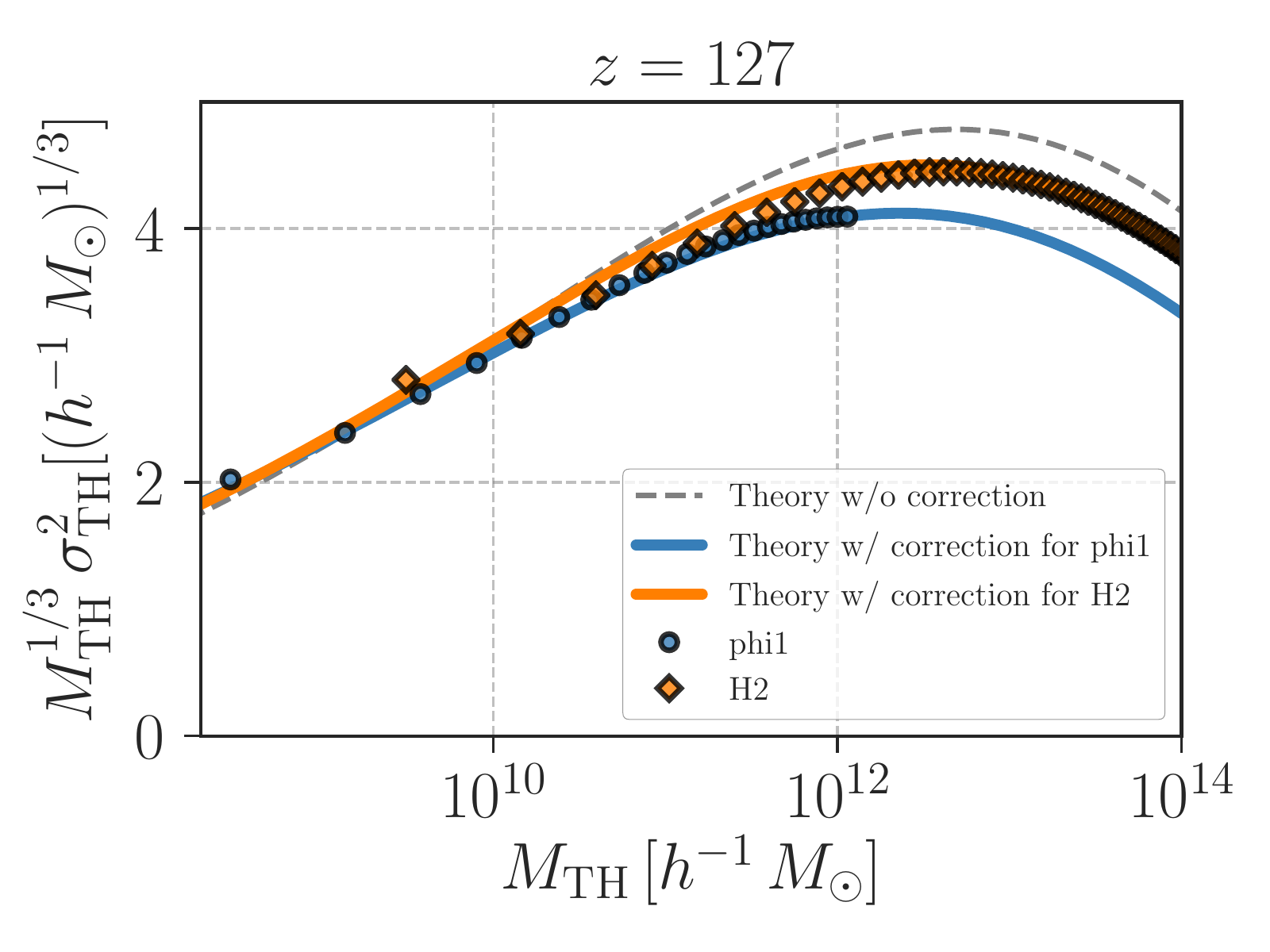}{0.5\textwidth}{}}
\caption{
Finite-volume effects in computing the linear top-hat mass variance $\sigma_\mathrm{TH}$.
The gray dashed line shows the prediction by linear perturbation theory (Eq.~\ref{eq:sigma_TH}), 
while the blue circles and orange diamonds are the mass variance measured in $N$-box boxes of phi1  ($L_\mathrm{box} = 32\, h^{-1}\mathrm{Mpc}$)
and H2 runs ($L_\mathrm{box} = 70\, h^{-1}\mathrm{Mpc}$) at $z=127$. 
The blue and orange lines are the mass variances with a correction by Eq.~(\ref{eq:sigma_loc}).
\label{fig:sigma_TH}}
\end{figure}

In Eq~(\ref{eq:fsim}), we require the logarithmic derivative of $\log \sigma$ 
with respect to the halo mass $M$. 
This derivative is known to be affected by the size of simulation boxes 
\citep[e.g.][]{2007MNRAS.374....2R, 2007ApJ...671.1160L}, 
because Fourier modes with scales beyond the box size are missed in the simulations.
To account for the finite volume effect on the estimate of $f(\sigma)$,
we follow an approach in \citet{2007MNRAS.374....2R}.
The mass variance in a finite-volume simulation $\sigma_\mathrm{loc}$ is not equal to the global value in Eq.~(\ref{eq:sigma_TH}).
Nevertheless, we assume that the halo mass function in the finite-volume simulation,
$\left(\mathrm{d}n/\mathrm{d}M\right)_\mathrm{loc}$, can be written as
\beqa
\left(\frac{\mathrm{d}n}{\mathrm{d}M}\right)_\mathrm{loc}
= \frac{\mathrm{d}n}{\mathrm{d}M}\Biggr{|}_{\sigma=\sigma_\mathrm{loc}}. \label{eq:loc_global_dndm}
\eeqa
This ansatz is motivated by the consideration in \citet{2002MNRAS.329...61S}.
Here, we define the halo mass function in the finite-volume simulation as
\beqa
\left(\frac{\mathrm{d}n}{\mathrm{d}M}\right)_\mathrm{loc} = 
\frac{\bar{\rho}_\mathrm{m}}{M} \frac{\mathrm{d}\ln \sigma^{-1}_\mathrm{loc}}{\mathrm{d}M}\, f_\mathrm{loc}(\sigma_\mathrm{loc}). \label{eq:dndm_loc}
\eeqa
Using Eqs.~(\ref{eq:loc_global_dndm}) and (\ref{eq:dndm_loc}), 
we can predict the global mass function with $f(\sigma) = f_\mathrm{loc}(\sigma)$
once we calibrate the functional form of $f_\mathrm{loc}$ as a function of $\sigma_\mathrm{loc}$.
It would be worth noting that Eq~(\ref{eq:fsim}) provides an estimate of $f_\mathrm{loc}(\sigma_\mathrm{loc})$ (not $f(\sigma_\mathrm{loc})$) in practice.
To evaluate the $\sigma_\mathrm{loc}-M$ relation, we directly compute the variance of smoothed density fields at the initial conditions of our simulations as varying smoothing scale of $R=(3M/4\pi\bar{\rho}_\mathrm{m})^{1/3}$.
To do so, we grid $N$-body particles onto meshes with $512^3$
cells using the cloud-in-cell assignment scheme and apply the three-dimensional 
Fast Fourier Transform (FFT).
Figure~\ref{fig:sigma_TH} shows 
the finite-volume effect of the mass variance measured in the phi1 and H2 runs at $z=127$.
We find that the finite-volume effect can be approximated as
\beqa
\sigma_\mathrm{loc}(M) = \sigma(M) \left(\frac{M}{M_0}\right)^{\eta}, \label{eq:sigma_loc}
\eeqa
where $M_0 = 2\times 10^{9}\, h^{-1}M_\odot$ and $\eta = -0.02$ give a reasonable fit to the phi1 run,
while $M_0 = 1\times 10^{10}\, h^{-1}M_\odot$ and $\eta = -0.01$ can explain $\sigma_\mathrm{loc}$
in the H2 run.
For the L run, we assume no finite-volume effects on the mass variance and set $\sigma_\mathrm{loc}=\sigma$.

To estimate $f_\mathrm{sim}$, we first measure the comoving number density of halos with 160 logarithmic bins in the range of $M=[10^{8}, 10^{16}]$. The bin size $\Delta \log M$ is set to 0.05.
We then compute the multiplicity function using Eq.~(\ref{eq:fsim}) with $\sigma\rightarrow\sigma_\mathrm{loc}$ 
as in Eq~(\ref{eq:sigma_loc}).

\subsection{Statistical errors}\label{subsec:stat_error}

For the calibration of our model with numerical simulations, 
we need a robust estimate of statistical errors in halo mass functions.
In this paper, we adopt an analytic model of the sample variance of $\mathrm{d}n/\mathrm{d}M$ 
developed in \citet{2003ApJ...584..702H}.
Apart from a simple Poisson noise, 
the model takes into account the fluctuation of the number density of dark matter halos 
in a finite volume.

Assuming that the fluctuation in $\mathrm{d}n/\mathrm{d}M$ is caused by underlying linear density modes
at their scales comparable to the simulation box size, one finds
\beqa
\frac{\mathrm{Err}[\Delta n_\mathrm{bin}]}{\Delta n_\mathrm{bin}}
&=& 
\left(\frac{1}{\Delta n_\mathrm{bin} V_\mathrm{sim}} + S(M_\mathrm{bin}, z)\right)^{1/2}, \label{eq:stat_error} \\
S(M, z) &=& 
b^2_L(M,z) \int\,\frac{k^2\mathrm{d}k}{2\pi^2} \, W^2_\mathrm{TH}(kR_\mathrm{box}) \nonumber \\
&&
\quad \quad \quad \quad 
\quad \quad \quad \quad 
\times P_\mathrm{L}(k, z), \label{eq:sample_variance}
\eeqa
where 
$\mathrm{Err}[\Delta n_\mathrm{bin}]$ represents the statistical error in Eq.~(\ref{eq:nbin}),
$V_\mathrm{sim} = L^3_\mathrm{box}$, $R_\mathrm{box} = [3V_\mathrm{sim}/(4\pi)]^{1/3}$,
and $b_\mathrm{L}(M,z)$ is the linear bias at the halo mass of $M$ and redshift $z$.
The first term in the right hand side in Eq.~(\ref{eq:stat_error}) corresponds to the Poisson error,
while the term of $S(M,z)$ represents the sample variance.
To compute Eq.~(\ref{eq:sample_variance}), we use the model of $b_\mathrm{L}$ in \citet{2010ApJ...724..878T}.

\begin{figure}
\gridline{\fig{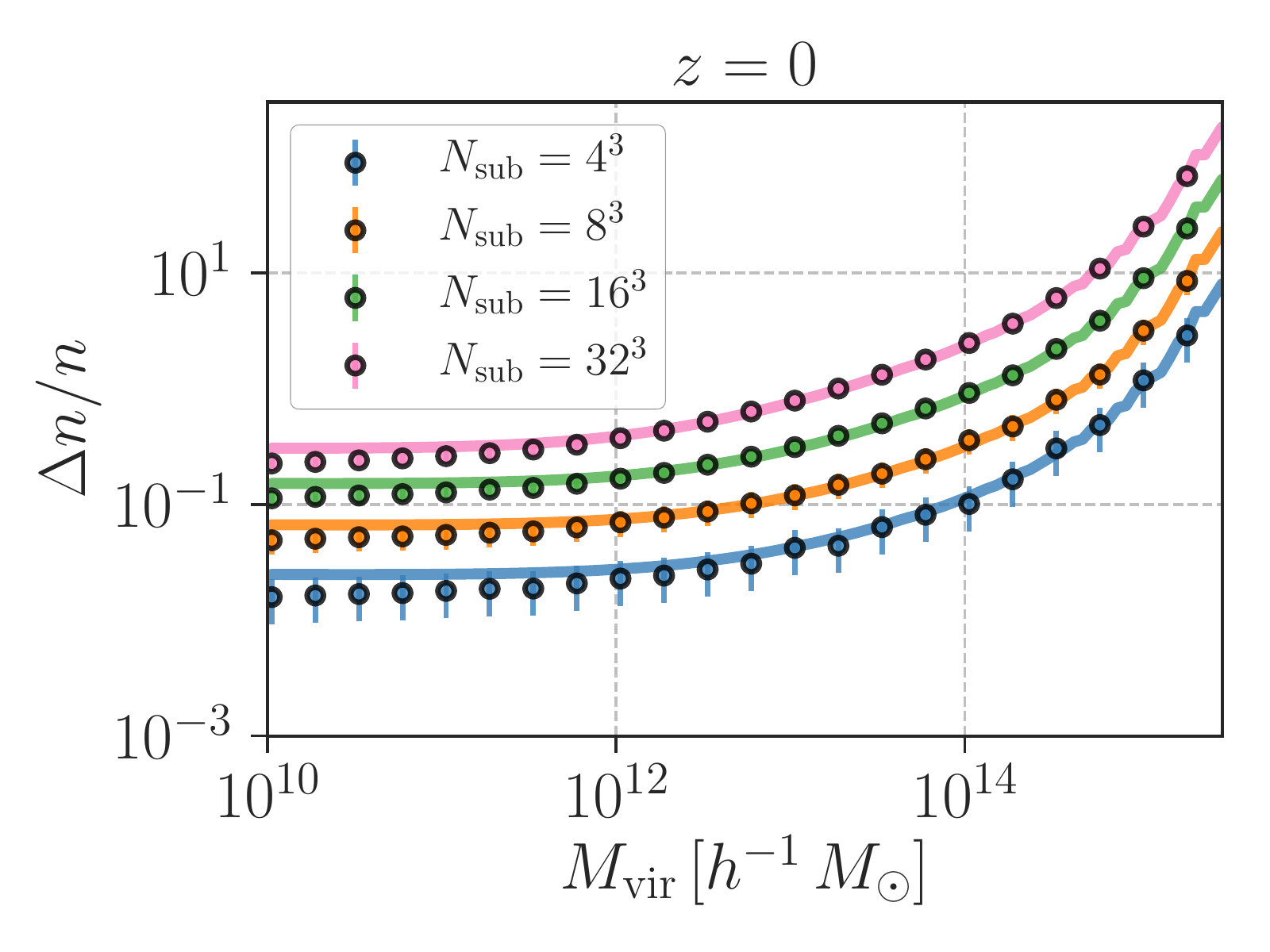}{.5\textwidth}{}}
\caption{
Sample variance of the mass function in a finite volume.
The colored points show the fractional error of the mass function in subvolumes in the L run at $z=0$, while the error bars represent the Gaussian error.
In this figure, we set the simulation results with the bin width being $\Delta \log M = 0.25$ 
for visualization.
From top to bottom, we show the results when dividing the L run into 
$N_\mathrm{sub} = 4^3, 8^3, 16^3$, and $32^3$ pieces, respectively.
Solid lines correspond to our model predictions based on Eq.~(\ref{eq:stat_error}),
showing a reasonable fit to the simulation results for different $N_\mathrm{sub}$.
\label{fig:sample_variance}}
\end{figure}

We validate the model of Eq.~(\ref{eq:stat_error}) with the L run at $z=0$.
In the validation, we divide the L run into $N_\mathrm{sub}$ subvolumes, 
compute the mass function in each subvolume, 
and then estimate the standard deviation of the mass function over the $N_\mathrm{sub}$ subvolumes.
We set each subvolume so that it has an equal volume.
Figure~\ref{fig:sample_variance} summarizes our validation.
In the figure, we show the fractional error of the binned mass function
and up-turns at higher masses indicate that the sample variance term becomes \ms{dominant}.
Because of Eq.~(\ref{eq:fsim}), the fractional error of $f_\mathrm{sim}$ is given by
Eq.~(\ref{eq:stat_error}).

\section{Calibration of the model parameters} \label{sec:calibration}

To calibrate our model of the multiplicity function (Eq.~\ref{eq:fmodel}) with the simulation,
we introduce a chi-square statistic at a given $z$ below:
\beqa
\chi^2_\mathrm{tot}(\bd{\theta}) &=& \chi^2_\mathrm{L}(\bd{\theta})
+ \chi^2_\mathrm{H2}(\bd{\theta}) + \chi^2_\mathrm{phi1}(\bd{\theta}), \label{eq:chisq_tot} \\
\chi^2_\alpha(\bd{\theta}) &\equiv& \sum_{i} \left(\frac{f_\mathrm{sim}(\sigma_i, \alpha)-f_\mathrm{mod}(\sigma_i, \bd{\theta})}{\mathrm{Err}_i + \sigma_{\mathrm{sys},i}}\right)^2,
\eeqa
where $f_\mathrm{sim}(\sigma_i, \alpha)$ represents 
the multiplicity function measured in the $\alpha$ run
($\alpha=\mathrm{L}, \mathrm{H2}, \mathrm{phi1}$) at the $i$-th bin of $\sigma$,
$f_\mathrm{mod}(\sigma_i, \bd{\theta})$ is given by Eq.~(\ref{eq:fmodel}) with the parameters being
$\bd{\theta}=\{\ln A(z),\ln a(z), \ln p(z), \ln q(z)\}$,
$\mathrm{Err}_i$ is the statistical error of the multiplicity function at $\sigma=\sigma_i$,
and $\sigma_{\mathrm{sys},i}$ is possible systematic errors at $\sigma=\sigma_i$ in our simulations
\ms{(we set $\sigma_{\mathrm{sys},i}$ later)}.
\ms{To find best-fit parameters, we minimize Eq.~(\ref{eq:chisq_tot})
with the Levenberg-Marquardt algorithm implemented in the open software of {\tt SciPy}\footnote{\url{https://www.scipy.org/}}\citep{scipy2001}.}

When computing Eq.~(\ref{eq:chisq_tot}), we reduce the number of bins in $\sigma$ by 
setting the bin width of $\Delta(1/\sigma) = 0.05$ for the L run 
and $\Delta(1/\sigma) = 0.1$ for other two runs.
This re-binning of $\sigma$ allows to reduce scatters among bins in the analysis and 
provide a reasonable goodness-of-fit for our best-fit model.
In our calibration, we do not include correlated scatters among bins \citep{2017MNRAS.469.4157C},
while we expect that our re-binning would make the correlation between nearest bins less important.
Also, we remove the data with $\Delta n_\mathrm{bin}$ being smaller than 30 
to avoid significant Poisson fluctuations in our fitting.
After these post-processes, we found $\sim40$ data points available at $z\le4.58$,
while $31$ and $25$ data points are left at $z=5.98$ and $7.00$, respectively.
Because our model consists of four parameters, we expect that our simulation data 
give sufficient information to find a good fit.

There exist several factors to introduce systematic effects 
on computing the mass function in $N$-body simulations \citep[e.g.][]{2005ApJS..160...28H, 2006MNRAS.373..369C, 2007ApJ...671.1160L, 2011MNRAS.415.2293K, 2013MNRAS.435.1618K, 2019MNRAS.488.3663L}.
Because we set the minimum halo mass to be $1000\, m_p$ with the particle mass of $m_p$ in the simulation,
the finite force resolution does not affect our measurement of the mass function 
beyond a $1\sigma$ Poisson error \citep{2019MNRAS.488.3663L}.
Although our simulations have a sufficient mass and force resolution and we properly correct the finite box effect in the measurement of mass functions (see Section~\ref{subsec:estimator}),
the identification of substructures in single halos can cause a systematic effect in our measurement.
There is no unique way to find substructures in $N$-body simulations \citep[e.g.][]{2012MNRAS.423.1200O},
and over-merging may take place even in the latest $N$-body simulations \citep[e.g.][]{2018MNRAS.475.4066V}.

Recently, \citet{2021ApJ...909..112D} found that the definition of boundary radii of host halos 
can change the subhalo abundance by a factor of $\sim2$, implying that our halo catalogs 
contain a non-negligible mislabeling of subhalos.
The mislabeled subhalos are mostly located at outskirts of host halos \citep{2021ApJ...909..112D}.
Note that we use the virial radius to identify subhalos in the simulation, but the virial radius does not correspond to a gravitational boundary radius in general \citep[e.g.][]{2015ApJ...810...36M, 2017ApJS..231....5D}.
Because more subhalos will be found as halo-centric radii increase, 
we expect that the difference of the multiplicity function with and without subhalos
can provide a reasonable estimate of systematic errors in our measurement.
In our simulation, we find that the multiplicity function with subhalos is different from one without subhalos in a systematic way.
The difference is less sensitive to the redshift and it can be well approximated as
\beq
\log \left(\frac{f_\mathrm{+s}(\sigma, z)}{f_\mathrm{fid}(\sigma, z)}\right)
= 
\left\{ \begin{array}{ll}
    0.05 & (\nu < 1) \\
    0.025\left(3-\nu\right) & (1\le \nu \le 3) \\
    0 & (\nu > 3)
  \end{array}
\right. , \label{eq:subhalo_sys}
\eeq
where 
$f_\mathrm{+s}$ is the multiplicity function with subhalos,
$f_\mathrm{fid}$ is the counterpart without subhalos (our fiducial data),
and $\nu = \delta_{c,z}/\sigma$.
Using Eq.~(\ref{eq:subhalo_sys}), we set $\sigma_\mathrm{sys}/f_\mathrm{fid} = \log (f_\mathrm{+s}/f_\mathrm{fid}) \times \ln 10$.
Note that our estimate of $\sigma_\mathrm{sys}$ should 
be overestimated because we assume that all subhalos are subject to mislabeling.
Hence, our analysis is surely conservative.


\section{Results} \label{sec:results}

\subsection{A simple check of non-universality}

\begin{figure}[t!]
\gridline{\fig{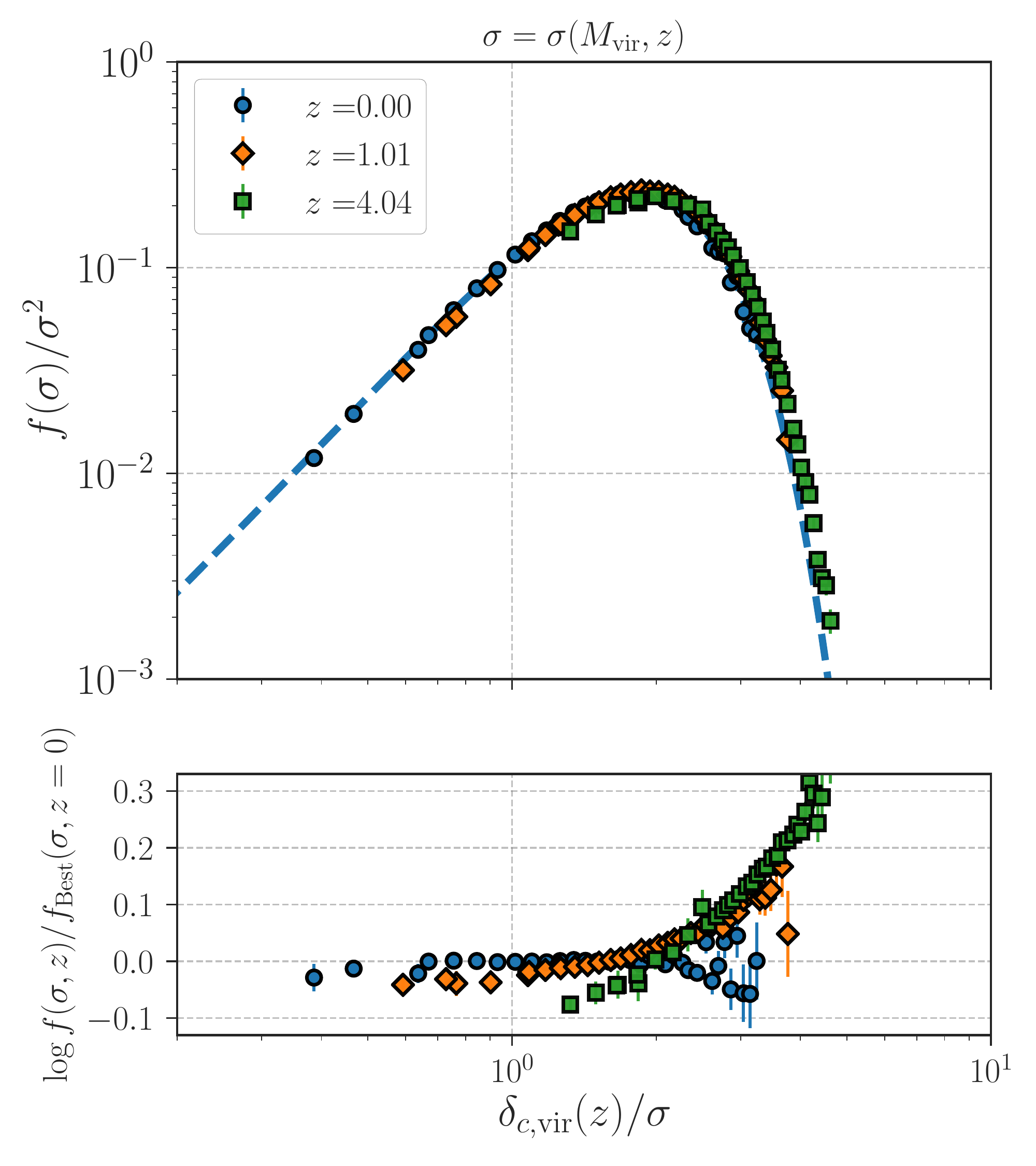}{0.5\textwidth}{}}
\caption{
Non-universality of virial halo mass functions in the $\nu^2$GC simulations.
The top panel shows the multiplicity function $f$ as a function of $\nu=\delta_{c,z}/\sigma$
at different redshifts. In the top, blue circles, orange diamonds, and green squares, represent
the simulation results at $z=0.00$, 1.01, and 4.04, respectively.
The blue dashed line in the top panel is the best-fit model of $f$ at $z=0.00$.
In the bottom, we show the difference of $\log f$ between the simulation results at the three redshifts
and the best-fit model at $z=0.00$.
If the multiplicity function is universal across redshifts, all the symbols should 
locate at $y=0$ in the bottom panel.
This figure clarifies that the virial halo mass function exhibits a strong redshift evolution.
\label{fig:fsigma_evol_simple}}
\end{figure}

Before showing the results of our calibration, we perform a sanity check to see the redshift dependence of the multiplicity function in the simulation.
Figure~\ref{fig:fsigma_evol_simple} shows the result of our sanity test.
In this figure, we first find a best-fit model for the multiplicity function at $z=0$, denoted as $f_\mathrm{Best}(\sigma, z=0)$.
We then compare the multiplicity function in the simulation at $z>0$ and $f_\mathrm{Best}(\sigma, z=0)$.
If the function $f(\sigma, z)$ is universal across different redshifts, we should find a residual between the simulation results at $z>0$ and $f_\mathrm{Best}(\sigma, z=0)$ as small as the case for $z=0$.
In the top panel in Figure~\ref{fig:fsigma_evol_simple}, the blue dashed line represents
the best-fit model $f_\mathrm{Best}(\sigma, z=0)$, while blue circles show the simulation result at $z=0$.
Comparing between the blue dashed line and blue circles, 
our fitting at $z=0$ provides a representative model of the simulation result.
The orange squares and green diamonds in the figure show the simulation results at $z=1.01$ 
and $z=4.04$, respectively. 
The bottom panel shows the residual between the simulation results and the model of $f_\mathrm{Best}(\sigma, z=0)$, highlighting prominent redshift evolution of 
the multiplicity function at $\delta_{c,z}/\sigma \simgt 3$ from $z=0$ to $4$.
In this figure, the error bars show the statistical uncertainties
and we account for the sample variance caused by finite box effects in our simulation (see Section~\ref{subsec:stat_error}).

\begin{figure*}
\gridline{\fig{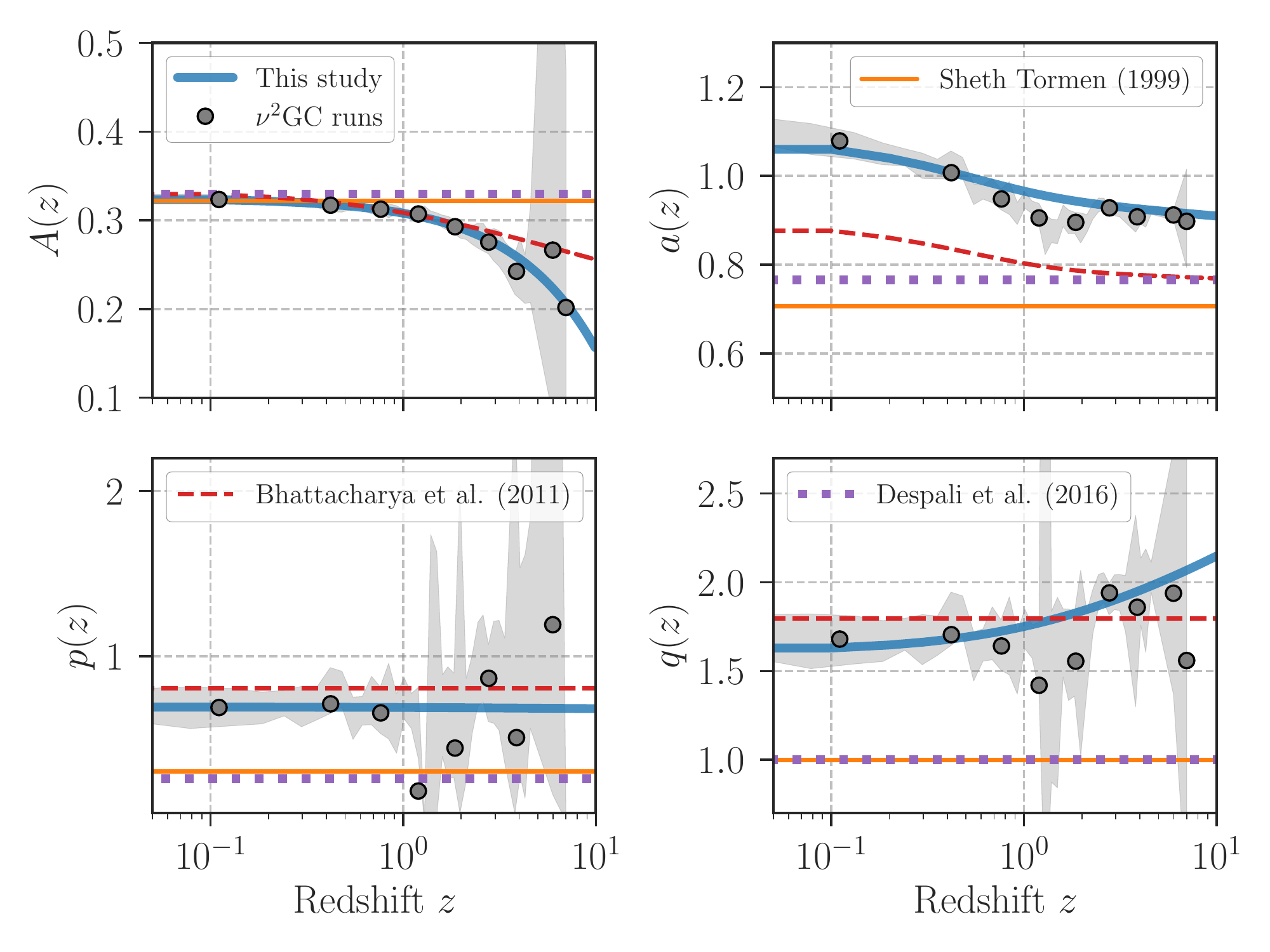}{1\textwidth}{}}
\caption{
Redshift evolution of parameters for virial halo mass functions in the $\nu^2$GC simulations.
In each panel, the shaded region shows $\pm1\sigma$ error bars inferred by our fitting, 
while the circles are best-fit parameters at representative coarse redshift bins (see the main text).
The blue line represents our calibrated model. 
For comparison, the red dashed line is the redshift-dependent model in \citet{2011ApJ...732..122B},
while the orange solid and purple dotted lines are the redshift-independent models in \citet{1999MNRAS.308..119S} and
\citet{2016MNRAS.456.2486D}, respectively.
\label{fig:bestfit_parameters}}
\end{figure*}

\subsection{Fitting results}

We here show the main result in this paper.
Figure~\ref{fig:bestfit_fsigma} in Appendix~\ref{apdx:fit_summary} summarizes the residual between the multiplicity functions in our simulations and the best-fit model at different redshifts.
We find that our fitting works well across a wide range of redshifts ($0\le z \le 7$).
A typical difference between the simulation results and our best-fit model is an of order of 0.02 dex except for
bins at $\delta_{c,z}/\sigma(z) \simgt 2-3$.
Although the bins at $\delta_{c,z}/\sigma(z) \simgt 2$ suffer from statistical fluctuations induced by the sample variance, we find that the residual is still 
within 0.06 dex even for such rare objects. 
The goodness-of-fit for our best-fit models ranges from 0.1 to 1.1.
It would be worth noting that we include possible systematic errors 
due to modeling of subhalos in our fitting.
Such systematic errors can reduce the score of $\chi^2$ for the best-fit model, making the best-fit $\chi^2$
smaller than the number of degrees of freedom.

The redshift evolution in best-fit parameters is shown in Figure~\ref{fig:bestfit_parameters}.
The gray shaded region in each panel shows $\pm1\sigma$ errors of a given parameter, inferred from the Jacobian of Eq.~(\ref{eq:chisq_tot}).
In each panel, the circles at $z\le4.57$ represent averages of the best-fit parameter within coarse bins of $z$, while the ones at $z=5.98$ and 7.00 show the best-fit parameters.
To compute the average, we set seven coarse redshift bins.
The edge of coarse bins is set to 
$[0.00,0.30)$, $[0.30,0.61)$, $[0.61,1.01)$,
$[1.01,1.49)$, $[1.49,2.44)$, $[2.44,3.36)$, and $[3.36,4.57]$.
We choose this binning of redshifts so that the dynamical time
(i.e. the ratio of the virial radius and virial circular velocity for halos) can be comparable to the Hubble time at bin-centered redshifts.

The blue solid lines in Figure~\ref{fig:bestfit_parameters} present our calibrated model. We find that the following form can explain the redshift evolution of the best-fit parameters and
return a similar level of the residual as in Figure~\ref{fig:bestfit_fsigma} when used in Eq.~(\ref{eq:fmodel}):
\beqa
A(z) &=& 0.325 - 0.017 \, z, \label{eq:model_Az}\\
a(z) &=& 0.940 \, (1+z)^{-0.02} \left(\frac{1.686}{\delta_{c,z}}\right)^2 \nonumber \\ 
&&
\qquad \times \left[1-0.015\left(\frac{z}{1.5}\right)^{0.01}\right]^{-1}, \label{eq:model_az} \\
p(z) &=& 0.692, \label{eq:model_pz} \\
q(z) &=& 1.611 \, (1+z)^{0.12}, \label{eq:model_qz}
\eeqa
where $\delta_{c,z}$ is given by Eq.~(\ref{eq:delta_cz}).

\subsection{Comparison with previous studies}

Our model (Eqs.~\ref{eq:model_Az}-\ref{eq:model_qz}) can be compared 
with previous models in the literature.
The most popular model in \citet{1999MNRAS.308..119S} predicts a universal multiplicity function 
$f(\sigma)$ with $A=0.322$, $a=0.707$, $p=0.3$ and $q=1.0$.
It would be worth noting that the parameter of $A$ in \citet{1999MNRAS.308..119S} is fixed by the normalization condition:
\beqa
\int_{-\infty}^{\infty}\, \mathrm{d}\ln\sigma \, f(\sigma) = 1, \label{eq:norm_f}
\eeqa
where it means that all dark matter particles reside in halos.
Because our model has been calibrated by the data with a finite range of $\sigma$,
it is not necessary to satisfy the condition of Eq.~(\ref{eq:norm_f}).
The integral of Eq.~(\ref{eq:norm_f}) for our model can be well approximated as $1.37\,(1+z/0.55)^{-0.35}\, (1-z/10.5)^{0.21}$ at $z\le 7$ within a 0.5\%-level accuracy.
In reality, the upper limit in the integral of Eq.~(\ref{eq:norm_f}) may be set by 
the free streaming scale of dark matter particles.
If dark matter consists of Weakly Interacting Massive Particle (WIMP) with a particle mass being $\sim100$ GeV,
the minimum halo mass is estimated to be $10^{-12}-10^{-3}\, M_{\odot}$ \citep[e.g.][]{2001PhRvD..64h3507H, 2003PhRvD..68j3003B, 2004MNRAS.353L..23G, 2005PhRvD..71j3520L, 2006PhRvD..74f3509B, 2006PhRvL..97c1301P, 2015PhRvD..92f5029D}.
When we set the upper limit to be $\sigma_\mathrm{upp}(z) = \sigma(M=10^{-6}\, h^{-1}M_{\odot}, z)$, the fraction of mass in field halos can be approximated as
\beqa
f_\mathrm{halo} &=& \int_{-\infty}^{\ln \sigma_\mathrm{upp}(z)}\, 
\mathrm{d}\ln\sigma \, f(\sigma, z) \nonumber \\
&\simeq& 0.724\,\left(1+\frac{z}{1.29}\right)^{-0.28}\, \left(1-\frac{z}{14.5}\right)^{1.02},
\label{eq:f_mass}
\eeqa
where the approximation is valid at $z\le7$ within a 0.4\%-level accuracy.
Eq.~(\ref{eq:f_mass}) is less sensitive to the choice of the minimum halo mass 
as long as we vary the minimum halo mass in the range of $10^{-12}-10^{-3}\, h^{-1}M_{\odot}$.
Our model predicts that about 72\% of the mass density in the present-day universe resides in dark matter halos.

For the parameter $a$, 
our model (Eq.~\ref{eq:model_az}) shows a modest redshift evolution with a level of $\sim20\%$
from $z=7$ to $z=0$.
Note that an effective critical density $\sqrt{a(z)}\, \delta_{c,z}$ becomes less dependent on $z$ and evolves only by $2-3$\% in the range of $0\le z \le 7$.
The redshift dependence of $a$ is mostly consistent with the model in \citet{2011ApJ...732..122B},
but the overall amplitude differs by $\sim20\%$.
Note that the model in \citet{2011ApJ...732..122B} has been calibrated for the halo mass function 
when the mass is defined by the Friend-of-friend (FoF) algorithm.
Because the FoF mass is expected to strongly depend on inner density profiles and substructures \citep{2011ApJS..195....4M}, our model needs not match the one in \citet{2011ApJ...732..122B}.
The present-day values of $a$, $p$, and $q$ are in good agreement with the result in \citet{2017MNRAS.469.4157C},
which presented the calibration of the halo mass function at $z=0$ 
with the MultiDark simulation \citep{2012MNRAS.423.3018P, 2016MNRAS.457.4340K}.
Note that \citet{2017MNRAS.469.4157C} adopted 
same halo finder, mass definition and cosmology as ours.

\citet{2016MNRAS.456.2486D} argued that the multiplicity function can be expressed as a universal function once one includes the redshift dependence of the spherical critical density $\delta_c$.
Although we adopt the redshift-dependent critical density as well, 
we find that the parameters in the multiplicity function depend on redshifts (also see Figure~\ref{fig:fsigma_evol_simple}).
The main difference between the analysis in \citet{2016MNRAS.456.2486D} and ours is 
halo identifications in $N$-body simulations.
We define halos by the FoF algorithm in six-dimensional phase-space, 
while \citet{2016MNRAS.456.2486D} adopted a spherical-overdensity algorithm 
with a smoothing scale being the distance to the tenth nearest neighbor 
for a given $N$-body particle.
Note that halo finders based on particle positions may not distinguish two merging halos.
The {\tt ROCKSTAR} algorithm utilizes the velocity information of $N$-body particles, allowing to very
efficiently determine particle-halo memberships even in major mergers.

\begin{table*}
\renewcommand{\thetable}{\arabic{table}}
\centering
\caption{Models of the multiplicity function $f(\sigma, z)$ selected in this paper.
Note that the functional form of $f$ is provided in Eq.~(\ref{eq:fmodel}) except for the models 
in \citet[][T08]{2008ApJ...688..709T} and \citet[][W13]{2013MNRAS.433.1230W}. The T08/W13 model assumes $f(\sigma,z)= A'\, [(\sigma/b')^{-a'}+1]\, \exp(-c'/\sigma^2)$ where $A', a', b'$ and $c'$ are redshift-dependent parameters. In this table, the second column shows which cosmological model is used to calibrate the functional form of $f$. WMAP1, WMAP3, WMAP5 and WMAP7 represent the first-, third-, fifth- and seventh-year WMAP constraints, respectively \citep{2003ApJS..148..175S, 2007ApJS..170..377S, 2009ApJS..180..330K, 2011ApJS..192...18K}. Planck14 means the best-fit cosmological model derived in \citet{2014A&A...571A..16P}, while Planck16 refers to the model in \citet{2016A&A...594A..13P}. The fourth column summarizes the mass definition of dark matter halos in simulations, while the fifth column shows if the parameters in $f$ are dependent on redshifts or not.
} \label{tab:mf_model}
\begin{tabular}{c|c|c|c|c}
\tablewidth{0pt}
\hline
\hline
Name
&
Cosmology
& 
Calibrated ranges of halo masses and redshifts
&
Halo mass
&
$z$-evolving $f$?
\\
\hline
\decimals
This paper
& Planck16
& $10^{8.5} \le M \, (h^{-1}M_\odot)\le 10^{15-0.45z}$ and $0 \le z \le 7$
& {\tt ROCKSTAR}
& Yes
\\
\hline
\decimals
\citet{1974ApJ...187..425P}
& --
& --
& --
& No
\\
\hline
\decimals
\citet{1999MNRAS.308..119S}
& --
& --
& --
& No
\\
\hline
\decimals
\citet{2008ApJ...688..709T}
& WMAP1, WMAP3
& $10^{11}\le M \, (h^{-1}M_\odot)\le 10^{15}$ and $0 \le z \le 2$
& SO
& Yes
\\
\hline
\decimals
\citet{2011ApJ...732..122B}
& WMAP5
& $6 \times 10^{11}\le M \, (M_\odot)\le 3\times 10^{15}$ and $0 \le z \le 2$
& FoF
& Yes
\\
\hline
\decimals
\citet{2013MNRAS.433.1230W}
& WMAP5
& $M \, (h^{-1}\, M_\odot) \ge 1.96\times 10^{9}$ at $z\simlt 8$
& SO
& Yes
\\
\decimals
& 
& $M \, (h^{-1}\, M_\odot) \ge 3.63\times 10^{6}$ at $8 \simlt z \le 26$
& 
& 
\\
\hline
\decimals
\citet{2016MNRAS.456.2486D}
& WMAP7, Planck14
& $5.8 \times 10^{9} \le M \, (h^{-1}M_\odot) \simlt 10^{15}$ and $0 \le z \le 5$
& SO
& No
\\
\hline
\end{tabular}
\end{table*}

\begin{figure*}
\gridline{
\fig{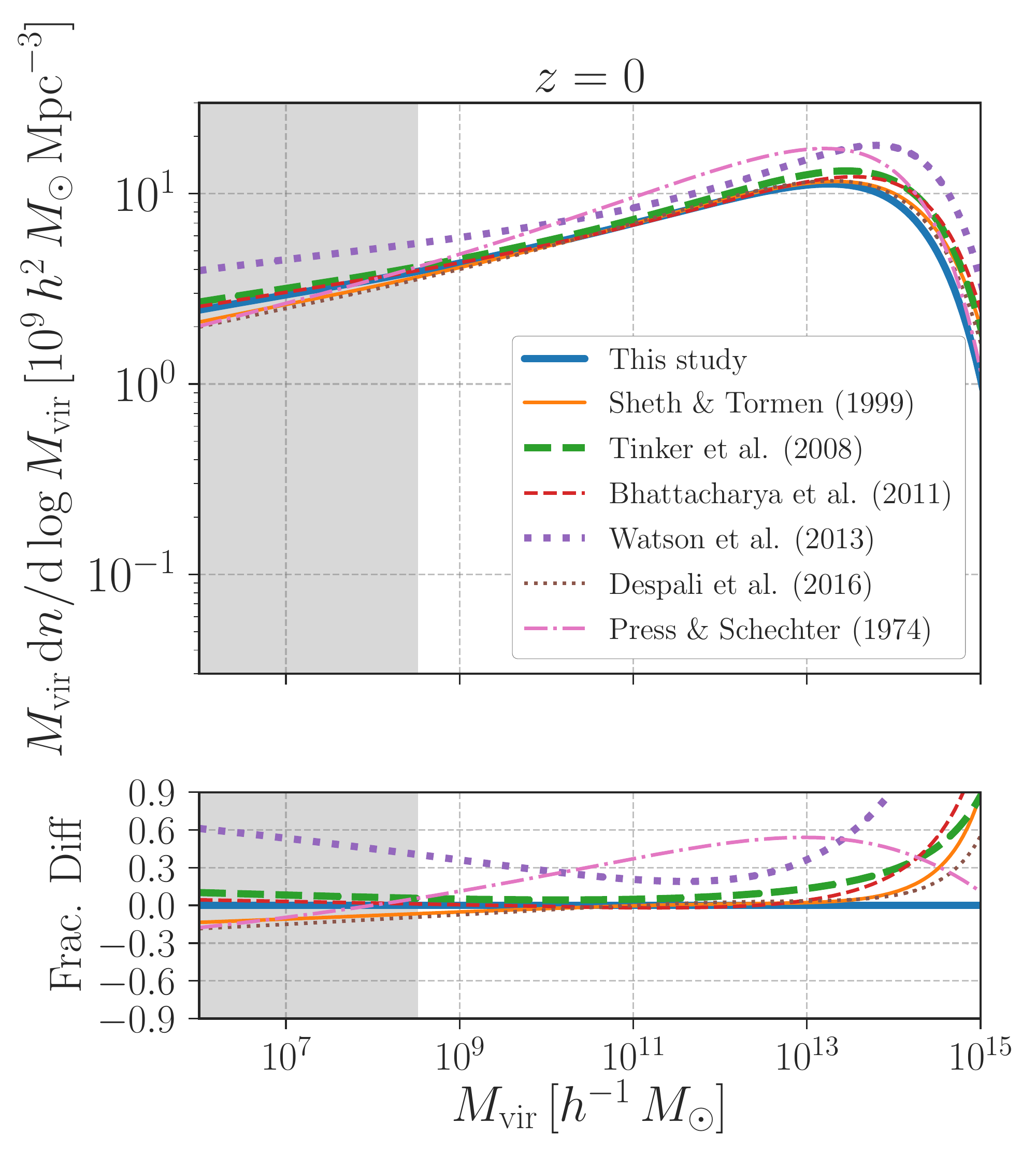}{0.5\textwidth}{}
\fig{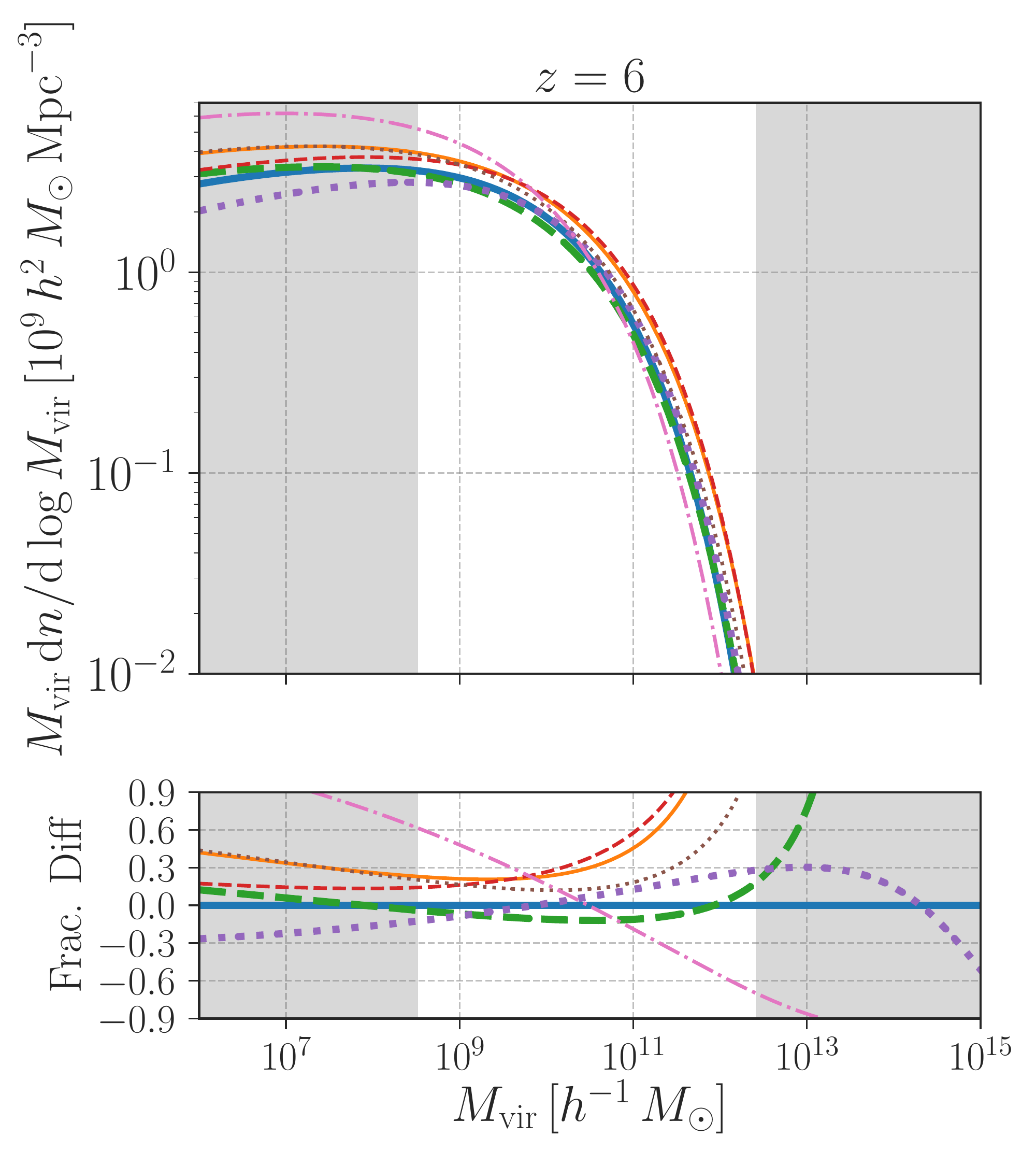}{0.5\textwidth}{}
}
\caption{
Comparison of fitting formulas for halo mass functions at $z=0$ and 6.
The left and right panels show the comparison at $z=0$ and 6, respectively.
Note that there are no available data for the gray shaded regime in our simulations.
In each panel, the blue thick solid line presents our result.
Other predictions in \citet{1974ApJ...187..425P}, \citet{1999MNRAS.308..119S}, \citet{2008ApJ...688..709T}, \citet{2011ApJ...732..122B}, \cite{2013MNRAS.433.1230W} and \citet{2016MNRAS.456.2486D} are also shown by pink dashed-dotted, orange thin solid, green thick dashed, red thin dashed, purple thick dotted, brown thin dotted lines, respectively.
\label{fig:comp_dndlogm}}
\end{figure*}

We next compare our model of the halo mass functions 
with previous models in the literature.
For comparison, we consider six representative models summarized in Table~\ref{tab:mf_model}.
Three of them assume a universal functional form of the multiplicity function, while others include some redshift evolution.
Among the previous studies, \citet{2016MNRAS.456.2486D} have investigated the mass function when using the virial halo mass, and their results can be directly compared to ours.
\citet{2008ApJ...688..709T} and \cite{2013MNRAS.433.1230W} studied mass functions for various spherical overdensity masses. We here use their fitting formula which takes into account the dependence of spherical overdensity parameters, 
while the calibration for the virial halo mass has not been done in \citet{2008ApJ...688..709T} and \cite{2013MNRAS.433.1230W}.
\citet{2011ApJ...732..122B} have calibrated the halo mass function when the mass is defined by the FoF algorithm, indicating that a direct comparison with our results may not be appropriate \citep[e.g. see][for details]{2011ApJS..195....4M}.

The left panel in Figure~\ref{fig:comp_dndlogm} shows 
the comparison of fitting formulas for the halo mass function at $z=0$,
while the right presents the case at $z=6$.
Our model is shown by blue solid lines in both panels
and the gray shaded region in the figure represents the range of halo masses 
not explored by our simulations.
At $z=0$, our model is in good agreement with most of previous models 
in the range of $10^{9-13}\, h^{-1}M_\odot$, 
but there are 15\%-level discrepancies at the high mass end ($M \simgt 10^{14} h^{-1}M_\odot$).
The figure implies that previous fitting formulas are not sufficient to predict
cosmology-dependence of the mass function for cluster-sized halos at $z=0$
even within a concordance $\Lambda$CDM cosmology.
A commonly-adopted model by \citet{2008ApJ...688..709T} may 
have a systematic error to predict the cluster abundance with a level of $20-30\%$, but the exact amount of systematic errors should depend on the definition of spherical overdensity halo masses.
We need larger $N$-body simulations \citep[e.g. see][for a relevant example]{2020arXiv200714720I}
to precisely calibrate the mass function at high mass ends 
and make a robust conclusion about systematic errors in cluster cosmology \citep[e.g.][]{2019ApJ...872...53M, 2020ApJ...901....5B, 2021MNRAS.504..769K}.
We leave further investigation of cluster mass functions for future studies.

At $z=6$, our model shows an offset from the universal models by \citet{1999MNRAS.308..119S} and \citet{2016MNRAS.456.2486D}.
The difference reaches a 15-20\% level at $M=10^{9-10}\, h^{-1}M_\odot$ 
and becomes larger at higher masses.
This is mainly caused by 
the redshift evolution of the amplitude in the multiplicity function, $A(z)$.
Our model predicts that the amplitude in the multiplicity function decreases as going higher redshifts, and this trend is clearly found in \ms{the simulation results} (see Figure~\ref{fig:fsigma_evol_simple}).

\begin{figure}
\gridline{\fig{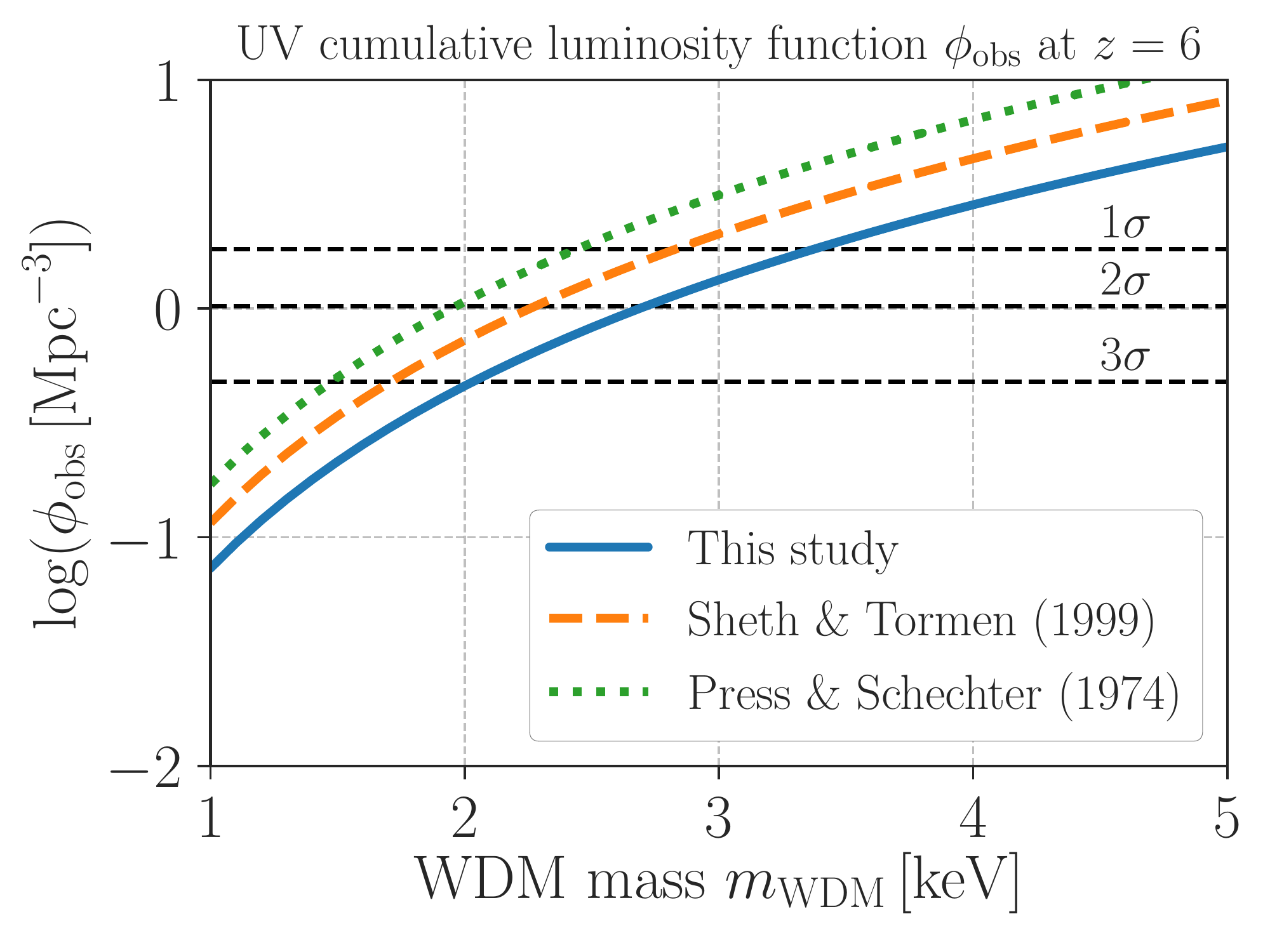}{.45\textwidth}{}}
\caption{
\ms{Lower} limits of warm dark matter (WDM) masses 
obtained from UV luminosity function of galaxies at $z=6$.
The dashed horizontal lines represent the lower bounds of the UV luminosity function at $z=6$ in the Hubble Frontier Fields as estimated in \citet{2016ApJ...825L...1M} ($1\sigma$, $2\sigma$ and $3\sigma$ levels from top to bottom). The solid line shows the maximum number density of dark matter halos as a function of WDM masses based on our calibrated halo mass functions and the correction proposed in \citet{2020ApJ...897..147L}. Assuming that an one-to-one correspondence of dark matter halos and faint galaxies, one can exclude the WDM model if the maximum number density of dark matter halos becomes smaller than the observed galaxy abundance. Our calibrated mass function places the lower limit of $m_\mathrm{WDM} > 2.71 \, \mathrm{keV}$ at the $2\sigma$ confidence level. This constraint is degraded with a level of 28\% and 16\% when one uses the commonly-adopted mass functions by \citet{1974ApJ...187..425P} and \citet{1999MNRAS.308..119S}, respectively.
\label{fig:UV_LF}}
\end{figure}

\subsection{Implications}\label{subsec:implications}

An important implication of our calibration is 
that modeling of the halo mass function for $\Lambda$CDM cosmologies can affect constraints of nature of dark matter particles by high-redshift galaxy number counts \citep[e.g.][]{2013MNRAS.435L..53P, 2014MNRAS.442.1597S, 2016ApJ...825L...1M, 2017PhRvD..95h3512C}.
Warm dark matter (WDM) is an alternative candidate of cosmic dark matter 
with free streaming due to its thermal motion.
Some physically-motivated extensions of the Standard model predict the existence of WDM 
with masses in keV range, such as sterile neutrino \citep[e.g.][]{2017JCAP...01..025A, 2019PrPNP.104....1B}.
Structure formation with WDM particles can be suppressed at scales below free-streaming lengths $\lambda_\mathrm{fs}$, while the bottom-up formation of dark matter halos remains as in the standard CDM paradigm at scales larger than $\lambda_\mathrm{fs}$.
A characteristic mass scale for $\lambda_\mathrm{fs}$ has been estimated as 
$\sim10^{10}\, M_\odot$ for WDM with a particle mass of $O(1)$ keV \citep[e.g.][]{2001ApJ...556...93B}.

In a hierarchical structure formation, less massive galaxies form at higher redshifts.
\ms{At a fixed cosmic mean density, WDM particles with larger masses 
are less effective at suppressing the growth of low-mass halos \citep[e.g.][]{2012MNRAS.424..684S}. 
Assuming that high-$z$ galaxies only form 
in collapsed halos, the observed abundance of high-$z$ galaxies can 
thus provide a lower limit to the particle mass of WDM.}

Recent high-resolution $N$-body simulations for WDM have indicated that the correction of the halo abundance due to free streaming can be expressed as a universal form of:
\beqa
\frac{\mathrm{d}n}{\mathrm{d}M}(M,z) \Biggr{|}_\mathrm{WDM} &=& {\cal R}(M)\, \frac{\mathrm{d}n}{\mathrm{d}M}(M,z) \Biggr{|}_\mathrm{CDM} \label{eq:dndM_WDM}\\
{\cal R}(M) &=& \left[1+\left(\frac{\alpha M_\mathrm{hm}}{M}\right)^{\beta}\right]^{\gamma},
\label{eq:correc_WDM}
\eeqa
where $\mathrm{d}n/\mathrm{d}M|_\mathrm{WDM}$ is the halo mass function for WDM cosmologies,
$\mathrm{d}n/\mathrm{d}M|_\mathrm{CDM}$ is the counterpart of CDM,
\citet{2020ApJ...897..147L} found that $\alpha=2.3$, $\beta=0.8$ and $\gamma=-1.0$ provide
a reasonable fit to simulation results.
In Eq.~(\ref{eq:correc_WDM}), $M_\mathrm{hm}$ is so-called ``half-mode" mass, 
which is defined as the mass scale that corresponds to the power spectrum wave number at which the square root of the ratio of the WDM and CDM power spectra is 0.5 \citep{2012MNRAS.424..684S}.
The mass $M_\mathrm{hm}$ depends on the particle mass of WDM $m_\mathrm{WDM}$:
\beqa
M_\mathrm{hm} &=& \frac{4\pi}{3}\, \bar{\rho}_m\, \left[s_\mathrm{WDM} \left(2^{\mu/5}-1\right)^{-\frac{1}{2\mu}}\right]^3, \\
s_\mathrm{WDM} &=& 0.153\, [h^{-1}\mathrm{Mpc}]\, \left(\frac{\Omega_\mathrm{WDM}}{0.25}\right)^{0.11} \nonumber \\
&&
\quad
\times \left(\frac{m_\mathrm{WDM}}{1\, \mathrm{keV}}\right)^{-1.11}
\left(\frac{h}{0.7}\right)^{1.22},
\eeqa
where $\mu=1.12$ and $\Omega_\mathrm{WDM}$ is the dimensionless density parameter of WDM.
According to Eq.~(\ref{eq:dndM_WDM}), an accurate calibration of mass function for $\Lambda$CDM cosmologies is essential to predict the counterpart of WDM.
\ms{We here caution that Eqs.~(\ref{eq:dndM_WDM}) and (\ref{eq:correc_WDM}) have been validated at $z=0$ 
and $z=2$ in \citet{2020ApJ...897..147L}. We assume that these equations are valid at $z\sim6$.
We leave a validation of Eqs.~(\ref{eq:dndM_WDM}) and (\ref{eq:correc_WDM}) at $z\simgt2$ for future studies.}

Figure~\ref{fig:UV_LF} demonstrates the importance of the calibration of CDM halo mass function 
when one constrains WDM masses with high-redshift galaxy number counts.
For given limiting magnitude and redshift, the cumulative galaxy number density 
should be smaller than the whole halo mass function within WDM cosmologies.
This leads 
\beqa
\phi_\mathrm{obs} \equiv
\int^{\infty}_{L_\mathrm{cut}} \mathrm{d}L\, \frac{\mathrm{d}n_\mathrm{gal}}{\mathrm{d}L} 
\le
\int^{M_\mathrm{max}}_{M_\mathrm{min}}\, \mathrm{d}M\, \frac{\mathrm{d}n}{\mathrm{d}M}\Biggr{|}_\mathrm{WDM}, \label{eq:get_WDM_mass}
\eeqa
where $L_\mathrm{cut}$ is the luminosity corresponding to the limiting magnitude,
$\mathrm{d}n_\mathrm{gal}/\mathrm{d}L$ is the galaxy luminosity function,
and we set $M_\mathrm{min}=1\, h^{-1}M_\odot$ and $M_\mathrm{max}=10^{16}\, h^{-1}M_\odot$.
Note that our lowest mass $M_\mathrm{min}=1\, h^{-1}M_\odot$ is much smaller than the half-mode mass $M_\mathrm{hm} \simgt 10^{10}\, h^{-1}M_\odot$ for $m_\mathrm{WDM}\simgt 1\, \mathrm{keV}$.
For the UV luminosity function at $z=6$ in the Hubble Frontier Fields with the limiting \ms{AB}
magnitude of $-12.5$ \citep{2017ApJ...835..113L}, the lower bound of $\phi_\mathrm{obs}$ has been estimated as
$\log (\phi_\mathrm{obs}\, [\mathrm{Mpc}^{-3}]) > 0.01$
at a $2\sigma$ confidence level \citep{2016ApJ...825L...1M}.
Assuming the best-fit cosmological parameters in \citet{2016A&A...594A..13P} and 
WDM is made of the whole abundance of dark matter, our model of the halo mass function with Eq.~(\ref{eq:correc_WDM}) allows to 
reject WDM with their mass smaller than 2.71 keV at the $2\sigma$ level.
This \ms{lower} limit is degraded to be 2.27 keV and 1.96 keV for the commonly-adopted models 
by \citet{1999MNRAS.308..119S} and \citet{1974ApJ...187..425P}, respectively.
This simple example highlights that 
\ms{calibration of the mass function for $\Lambda$CDM cosmologies is
essential to accurate predictions of the mass function for WDM 
cosmologies}.

\ms{Our lower limit of the WDM particle mass can be compared to other methods.
For example, \citet{2016JCAP...08..012B} found a 
lower limit of 2.96 keV from observations of the Lyman-alpha forest, 
while \citet{2020JCAP...04..038P} placed a lower limit of 5.3 keV 
(a similar limit is found by \citet{2017PhRvD..96b3522I}). 
\citet{2019MNRAS.487.3560C} used high-redshift 21-cm data from EDGES to rule out WDM 
with $m_\mathrm{WDM} < 3 \, \mathrm{keV}$.
Note that our limit is less sensitive to details of baryonic physics than the others, 
because our analysis relies on the cumulative abundance of dark matter halos.}

\ms{For a conservative analysis, we consider that a significant small halos of $M=1\, h^{-1}M_\odot$ 
can be responsible to the observed galaxy abundance at high redshifts.}
\ms{To further tighten the limit of WDM particle mass, 
it would be interesting to discuss more realistic halo mass scales to the faintest galaxy at $z\sim6$.}
\ms{In the Planck $\Lambda$CDM cosmology, we find that the minimum halo mass of $\sim10^{7}\, h^{-1}M_\odot$ provides the cumulative halo abundance of $1.35\, \mathrm{Mpc}^{-3}$, 
which is close to the observed galaxy abudnance at $z=6$.}
\ms{When setting the minimum halo mass to $10^{7}\, h^{-1}M_\odot$ in Eq.~(\ref{eq:get_WDM_mass}),
we find a stringent 2$\sigma$ limit of $m_\mathrm{WDM} > 14.1\, \mathrm{keV}$.}
\ms{However, this stringent limit is very sensitive to the choice of the minimum halo mass.
For the minimum halo mass of $10^{6}\, h^{-1}M_\odot$, the limit changes to $m_\mathrm{WDM} > 6.23\, \mathrm{keV}$.}
\ms{This simple analysis implies that a more detailed modeling of galaxy-halo connections at $M_\mathrm{vir}\sim10^{6-7}\, h^{-1}M_\odot$ and $z\sim6$ would be worth pursuing in future work.}


\section{Limitations} \label{sec:limitations}

We summarize the major limitations in our model of virial halo mass functions.  
All of the following issues will be addressed in forthcoming studies.

\subsection{Cosmological dependence}

\begin{figure}[t!]
\gridline{\fig{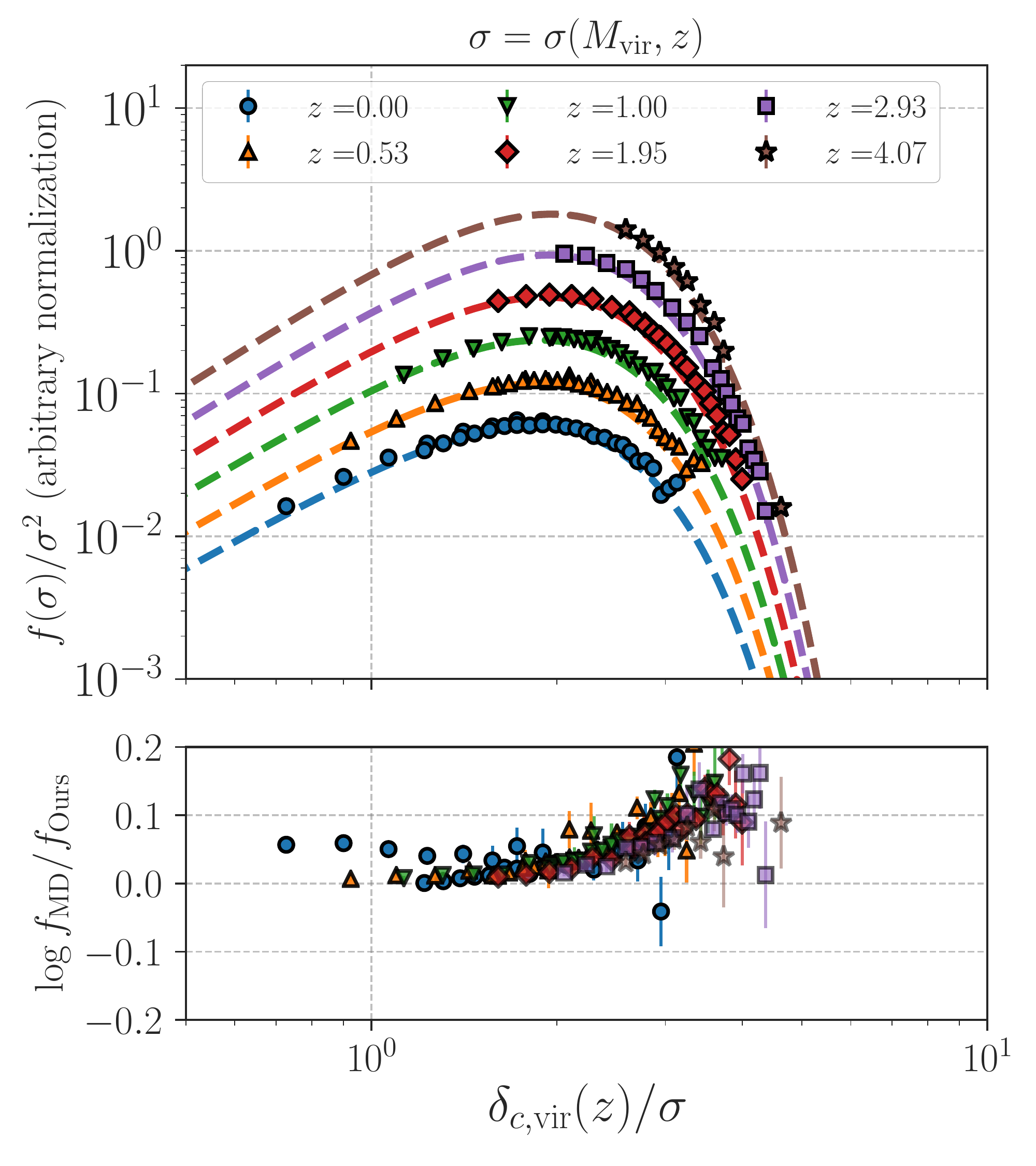}{.45\textwidth}{}}
\caption{
Comparison of the multiplicity function $f$ measured in the MultiDark-Run1 and Bolshoi (MB) simulations \citep{2011arXiv1109.0003R, 2011ApJ...740..102K, 2012MNRAS.423.3018P} 
and predictions by our fitting formula. 
Different symbols in each panel show the simulation results at various redshifts.
The dashed lines in the top panel show our predictions.
Note that we introduce an arbitrary shift in $f/\sigma^2$ for visibility in the top panel.
In the bottom, we show the difference of $\log f$ between the simulation results and our predictions.
Note that the MB simulations adopt the cosmological model with $\Omega_\mathrm{m0}=0.27$, while our fitting formula of $f$ relies on the simulation with $\Omega_\mathrm{m0}=0.31$. Other $\Lambda$CDM parameters are almost same between the two.
The bottom panel shows that our fitting formula of $f$ can have a systematic error with a level of 0.05-0.2 dex if one changes $\Omega_\mathrm{m0}$ by 13\%.
\label{fig:fsigma_evol_MD}}
\end{figure}

Our model of halo mass functions is calibrated against 
$N$-body simulations in the $\Lambda$CDM cosmology consistent with Planck16.
In terms of studies of large-scale structure, 
$\Omega_{\mathrm{m0}}$ and $\sigma_8$ are the primary parameters and 
the simulations in this paper adopt $\Omega_{\mathrm{m0}}=0.31$ and $\sigma_8=0.83$.
Therefore, our calibration of Eqs.~(\ref{eq:model_Az})-(\ref{eq:model_qz}) 
may be subject to an overfitting to the specific cosmological model.
To examine the dependence of our model on cosmological models, 
we use another halo catalog from $N$-body simulations with a different $\Lambda$CDM model.
For this purpose, we use the Bolshoi simulation in \citet{2011ApJ...740..102K} and the first MultiDark-Run1 simulation performed in \citet{2012MNRAS.423.3018P}.
The Bolshoi simulation consists of $2048^3$ particles in a volume of $250^3\, [h^{-1}\mathrm{Mpc}]^3$ and assumes the cosmological parameters of
$\Omega_{\mathrm{m0}}= 0.27$,
$\Omega_{\mathrm{b0}}= 0.0469$,
$\Omega_{\Lambda}=1-\Omega_{\mathrm{m0}} = 0.73$,
$h= 0.70$, $n_s= 0.95$, and $\sigma_8= 0.82$.
These are consistent with the five-year observation of the cosmic microwave background obtained by the WMAP satellite \citep{2009ApJS..180..330K}
and we refer to them as the WMAP5 cosmology.
The MultiDark-Run1 simulation adopted the same cosmological model
and the number of particles as in the Bolshoi simulation, while the volume is set to $1\, [h^{-1}\mathrm{Gpc}]^3$.
We use the {\tt ROCKSTAR} halo catalog at $z=0, 0.53, 1.00, 1.96, 2.93$ and $4.07$ from 
the Bolshoi and MultiDark-Run1 simulations.\footnote{The halo catalogs at different redshifts are publicly available at \url{https://slac.stanford.edu/~behroozi/MultiDark_Hlists_Rockstar/} and
\url{https://www.slac.stanford.edu/~behroozi/Bolshoi_Catalogs/}.}
To compute our model prediction for the WMAP5 cosmology,
we fix the functional form of Eq.~(\ref{eq:fmodel}) and parameters in Eqs.~(\ref{eq:model_Az})-(\ref{eq:model_qz}) but include the cosmology-dependence of 
the critical density $\delta_{c,z}$ and the top-hat mass variance $\sigma$, accordingly.

Figure~\ref{fig:fsigma_evol_MD} summarizes the multiplicity function in the Bolshoi and MultiDark-Run1 simulations.
In this figure, dashed lines show the predictions by our model for the WMAP5 cosmology, while
different symbols represent the simulation results.
We find that our model can reproduce simulation results within a 10\% level even for the WMAP5 cosmology at $0.7 \simlt \delta_{c,z}/\sigma \simlt 2.5$.
At high mass ends ($\delta_{c,z}/\sigma \simgt 2.5$), our model tends to underestimate the halo abundance by $\sim30-40\%$ in a wide range of redshifts.
It is worth noting that the residual between our model and the WMAP5-based simulation
is less dependent on redshifts.
In fact, we found a better matching between our models and the WMAP5-based simulations
when reducing the overall amplitude in the parameter $a(z)$ by $4-5\%$.
In summary, our model can not predict the simulation results for the WMAP5 cosmology with the same level as in the Planck cosmology.
The 10\%-level difference in $\Omega_{\mathrm{m0}}$ can 
cause systematic uncertainties in our model predictions with a level of 10\% except for high mass ends.
\ms{Future studies would need to calibrate the cosmological dependence of $a(z)$ for precision cosmology based on galaxy clusters.}
\ms{At low masses and high redshifts, our model can provide a reasonable fit to the simulations adopting the WMAP5 cosmology. According to this fact, we examine how much the WDM limit in Section~\ref{subsec:implications} is affected by the choice of underlying cosmology. For the WMAP5 cosmology, we find a 2$\sigma$ limit of $m_\mathrm{WDM} > 2.75\, \mathrm{keV}$, which differs from our fiducial limit by only $\sim1.4\%$.}

\subsection{Baryonic effects}

Our calibration of halo mass functions relies on dark-matter-only $N$-body simulations 
and ignores possible baryonic effects.
Baryonic effects on halo mass functions have been studied 
with a set of hydrodynamical simulations \citep[e.g.][]{2009MNRAS.394L..11S, 2012MNRAS.423.2279C, 2013MNRAS.431.1366S, 2014MNRAS.441.1769C, 2014MNRAS.439.2485C, 2014MNRAS.440.2290M, 2016MNRAS.456.2361B, 2021arXiv210305076B}.
The evolution of cosmic baryons is governed not only by gravity but also complex processes associated with galaxy formation. 
Relevant processes include gas cooling, star formation and 
energy feedback from supernovae (SN) and Active Galactic Nuclei (AGN).
Adiabatic gas heated only by gravitational processes affects the halo mass function with a level of $\simlt 2-3\%$ \citep{2012MNRAS.423.2279C}, while radiative cooling, star formation and SN feedback increase individual halo masses due to condensation of baryonic mass at the halo center, 
changing the halo mass function \citep{2009MNRAS.394L..11S, 2012MNRAS.423.2279C}.
Efficient SN feedback can decrease the halo mass function at $M \simlt 10^{11}\, M_\odot$ by $\sim20-30\%$ \citep{2013MNRAS.431.1366S}.
The mass function at $M \simeq 10^{13-14}\, M_\odot$ would be affected by AGN feedback, 
while current hydrodynamical simulations adopt a sub-grid model to include the AGN feedback.
Because a variety of sub-grid models has been proposed, the impact of the AGN feedback on the mass function is still uncertain \citep{2014MNRAS.441.1769C, 2014MNRAS.439.2485C, 2014MNRAS.440.2290M, 2016MNRAS.456.2361B}.

To account for the baryonic effects on the halo mass function, 
it is important to correct individual halo masses according to baryonic processes.
Baryons do not change the abundance of dark matter halos, but affect internal structures
and spherical masses of halos.
Hence, one may be able to model the halo mass function in the presence of baryons 
by abundance matching between the gravity-only and hydrodyamical simulations for a given definition of the halo mass \citep[e.g][]{2021arXiv210305076B}.
In this sense, our calibration of the halo mass function 
provides a baseline model and still plays an important role in understanding 
the baryonic effects on large-scale structures.

\section{Discussion and Conclusion} \label{sec:conclusion}

In this paper, we have studied mass functions 
in the concordance $\Lambda$ cold dark matter ($\Lambda$CDM) model 
inferred from the measurement of cosmic microwave backgrounds 
by the Planck satellite \citep{2016A&A...594A..13P}.
We have calibrated the abundance of dark matter halos in a set of $N$-body simulations
\citep{2015PASJ...67...61I, 2020MNRAS.492.3662I} covering a wide range of redshifts and halo masses.
For a theoretically-motivated virial spherical over-density mass $M$, 
we have employed least-square analyses to find best-fit models to our simulation results
in the range of $10^{8.5} \le M\, [h^{-1}M_\odot] \simlt 10^{15-0.45z}$ 
where redshifts $z$ range from 0 to 7.

Our calibrated models are able to reproduce the simulation results with a 5\%-level precision
over all redshifts explored in this paper, but except for high-mass ends.
We found that the multiplicity function $f$ defined in Eq.~(\ref{eq:dndm}) 
exhibits some redshift dependence, contradicting the commonly-adopted analytic model 
as in \citet[][ST99]{1999MNRAS.308..119S}.
The redshift evolution of the multiplicity function is prominent in our simulation data 
even at high redshifts $z\simgt3$.
Our calibrated halo mass function is in good agreement 
with previous models in the literature within a level of $\simlt15\%$ at $z=0$,
while our model predicts that the halo mass function in the range 
of $M=10^{8.5-10}\, h^{-1}M_\odot$ at $z=6$ can be smaller than the ST99 prediction by $20-30\%$.

If cosmic dark matter consists of 
Weakly Interacting Massive Particle (WIMP) with a particle mass of $\sim100$ GeV, 
the minimum halo mass would be of an order of $10^{-12}-10^{-3}\, h^{-1}M_\odot$ \citep[e.g.][]{2001PhRvD..64h3507H, 2003PhRvD..68j3003B, 2004MNRAS.353L..23G, 2005PhRvD..71j3520L, 2006PhRvD..74f3509B, 2006PhRvL..97c1301P}.
An extrapolation of our calibrated halo mass function to such minimum halo masses allows us to predict the fraction of cosmic mass density containing in halos. 
We found that the fraction can be well approximated as $0.724\, (1+z/1.29)^{-0.28}\, (1-z/14.5)^{1.02}$ at $0\le z \le 7$.
This implies that about 72\% of the mass density at present is confined in 
gravitationally-bound objects.
If cosmic dark matter consists of warm dark matter (WDM) with a particle mass of $\sim1$ keV such as sterile neutrino \citep[e.g.][]{2017JCAP...01..025A, 2019PrPNP.104....1B},
our model with a recently-proposed WDM correction \citep{2020ApJ...897..147L} 
provides a powerful test of WDM scenarios by comparing with galaxy number counts at high redshifts.
We found that WDM with a particle mass smaller than 2.71 keV is incompatible to the UV luminosity function at $z=6$ in the Hubble Frontier Fields \citep{2017ApJ...835..113L} at a $2\sigma$ confidence level.
It would be worth noting that this upper limit 
can be degraded by $16\%$ when one adopts the ST99 prediction.
This highlights that the calibration of halo mass functions in $\Lambda$CDM cosmologies
is required to place cosmological constraints of WDM especially 
when using high-redshift observables.

Our fitting formula of the virial halo mass function is based on dark-matter-only $N$-body simulations for a specific cosmology.
We found a 10\%-level difference in the average cosmic mass density 
can cause systematic uncertainties in our model predictions with a $\simlt 10\%$ level.
Baryonic effects such as gas cooling, star formation, and some feedback processes 
can affect internal structures of dark matter halos.
Although it is still difficult to account for the baryonic effects in our model,
recent simulations indicate that abundance matching between the gravity-only and hydrodynamical simulations would be promising \citep[e.g.][]{2021arXiv210305076B}.
This implies that our calibrated model is still meaningful as a baseline prediction before including baryonic effects, while more detailed analysis with hydrodynamical simulations is demanded.
Our analysis pipeline can be applied to any $N$-body simulations based on 
non-CDM, e.g. allowing to validate a universal suppression of the halo abundance (see Eq.~\ref{eq:correc_WDM}) in WDM cosmologies.

\acknowledgments
This work is in part supported by MEXT KAKENHI Grant Number (19K14767, 19KK0344, 20H05850, 20H05861, 21H01122).
Numerical computations were in part carried out on Cray XC50 at Center 
for Computational Astrophysics, National Astronomical Observatory of Japan.
TI has been supported by MEXT as
``Program for Promoting Researches on the Supercomputer Fugaku''
(Toward a unified view of the Universe: from large scale structures to
planets, proposal numbers hp200124 and hp210164), and JICFuS.
The $\nu^2$GC and Phi-1 simulations were run on the K computer 
at the RIKEN Advanced Institute for Computational Science and the Aterui 
supercomputer at Center for Computational Astrophysics, 
of National Astronomical Observatory of Japan.

\software{GreeM \citep{2009PASJ...61.1319I, 2012arXiv1211.4406I},
2LPTic \citep{2006MNRAS.373..369C},
MUSIC \citep{2011MNRAS.415.2101H},
CAMB \citep{Lewis:1999bs},
ROCKSTAR \citep{2013ApJ...762..109B},
Scipy \citep{scipy2001}}

\appendix
\section{Impact of unbound particles in halo mass definition}\label{apdx:strict_so_test}

\begin{figure}[t!]
\gridline{\fig{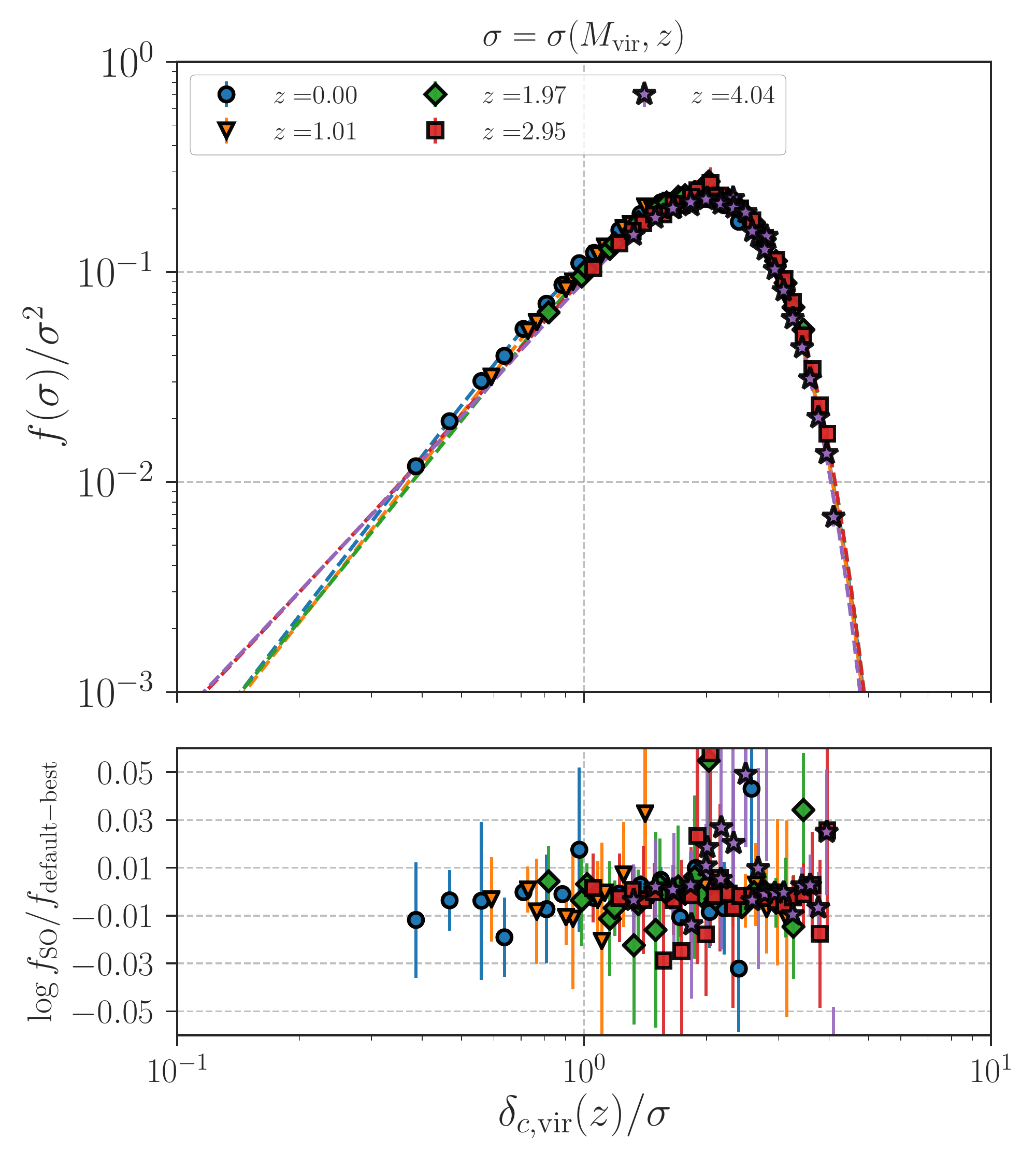}{.45\textwidth}{}}
\caption{
Comparison of the multiplicity function $f$ with and without unbound particles.
Different symbols in the top panel show the simulation results including unbound particles 
at various redshifts, while the dashed lines in the top panel show the best-fit models for our fiducial halo catalogs (without unbound particles).
In the bottom, we show the difference of $\log f$ between the two
and the error bars show statistical uncertainties.
\label{fig:fsigma_evol_SO}}
\end{figure}

We here examine possible effects of unbound particles around dark matter halos on
our calibration of the halo mass function.
For this purpose, we use $N$-body simulations with $4096^3$ particles and the box length on a side being $560\, h^{-1}\mathrm{Mpc}$, referred to as the $\nu^2$GC-M run in \citet{2015PASJ...67...61I}.
We prepare two different halo catalogs at redshifts of $z=0.00, 1.01, 1.97, 2.95$ and $4.04$.
One is the catalog with the default option for the {\tt ROCKSTAR} finder and does not include unbound particles in the virial mass for individual halos,
while another imposes the option of {\tt STRICT\_SO\_MASSES=1} to account for unbound particles.
Figure~\ref{fig:fsigma_evol_SO} compares the multiplicity functions measured in the two halo catalogs.
We find that our fitting of the multiplicity function with the default halo catalogs (our fiducial runs) is not affected by the inclusion of unbound particles beyond the statistical errors for the $\nu^2$GC-M runs.

\section{Summary of our fitting results}\label{apdx:fit_summary}

\begin{figure*}[t!]
\includegraphics[angle=270, width=0.9\linewidth]{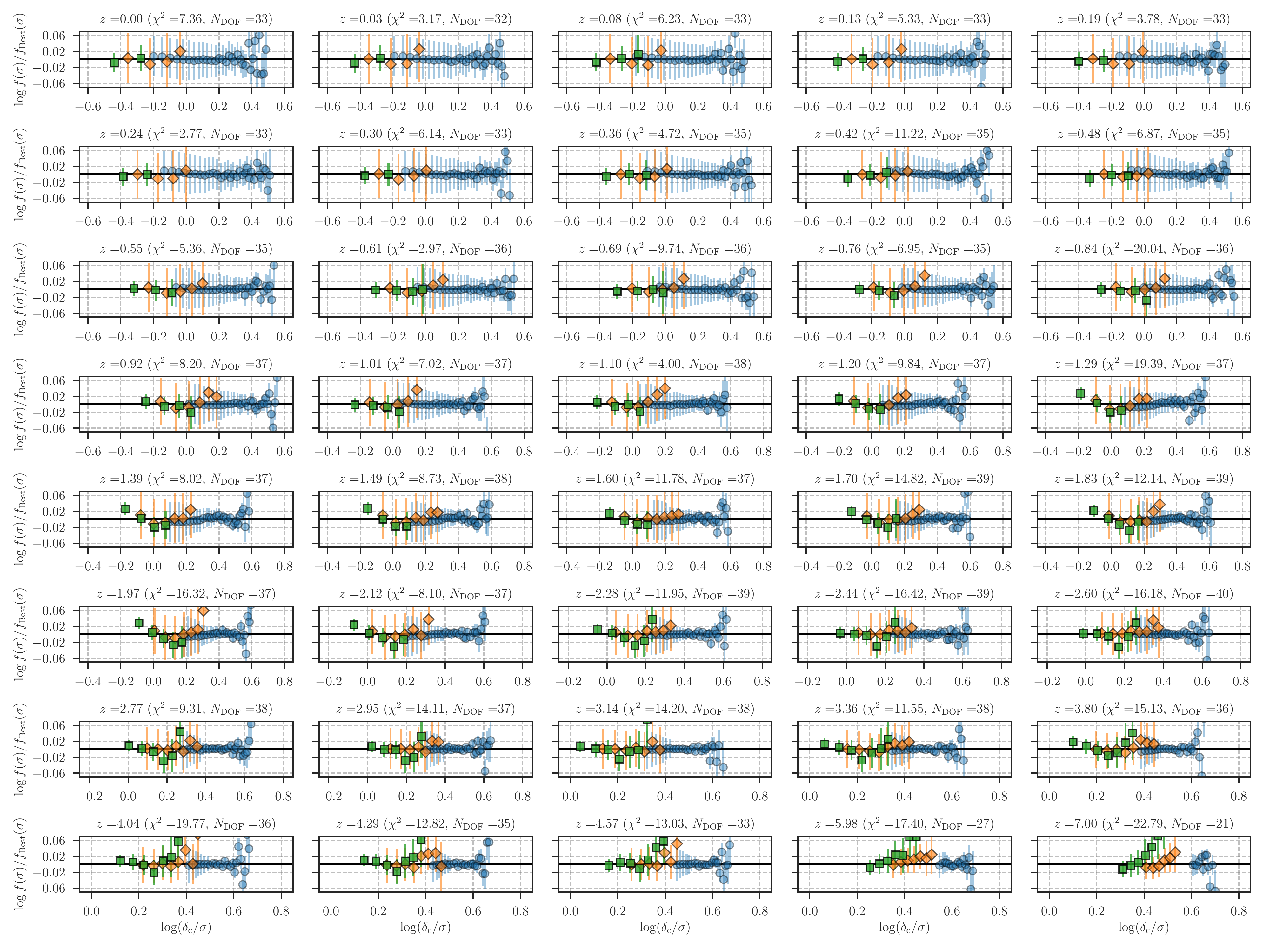}
\caption{
A summary of our fits to virial halo mass functions in the $\nu^2$GC simulations.
Each panel shows the residual of the multiplicity function $f$ in logarithmic space at different redshifts ($0\le z \le 7$). The colored symbols in each panel represent the residual between the simulation result and its best-fitted model and the black line shows no difference between the two.
The blue circles, orange diamonds, and green squares are the results in the L, H2,
and phi1 runs, respectively.
\label{fig:bestfit_fsigma}}
\end{figure*}

\begin{figure*}[t!]
\includegraphics[angle=270, width=0.9\linewidth]{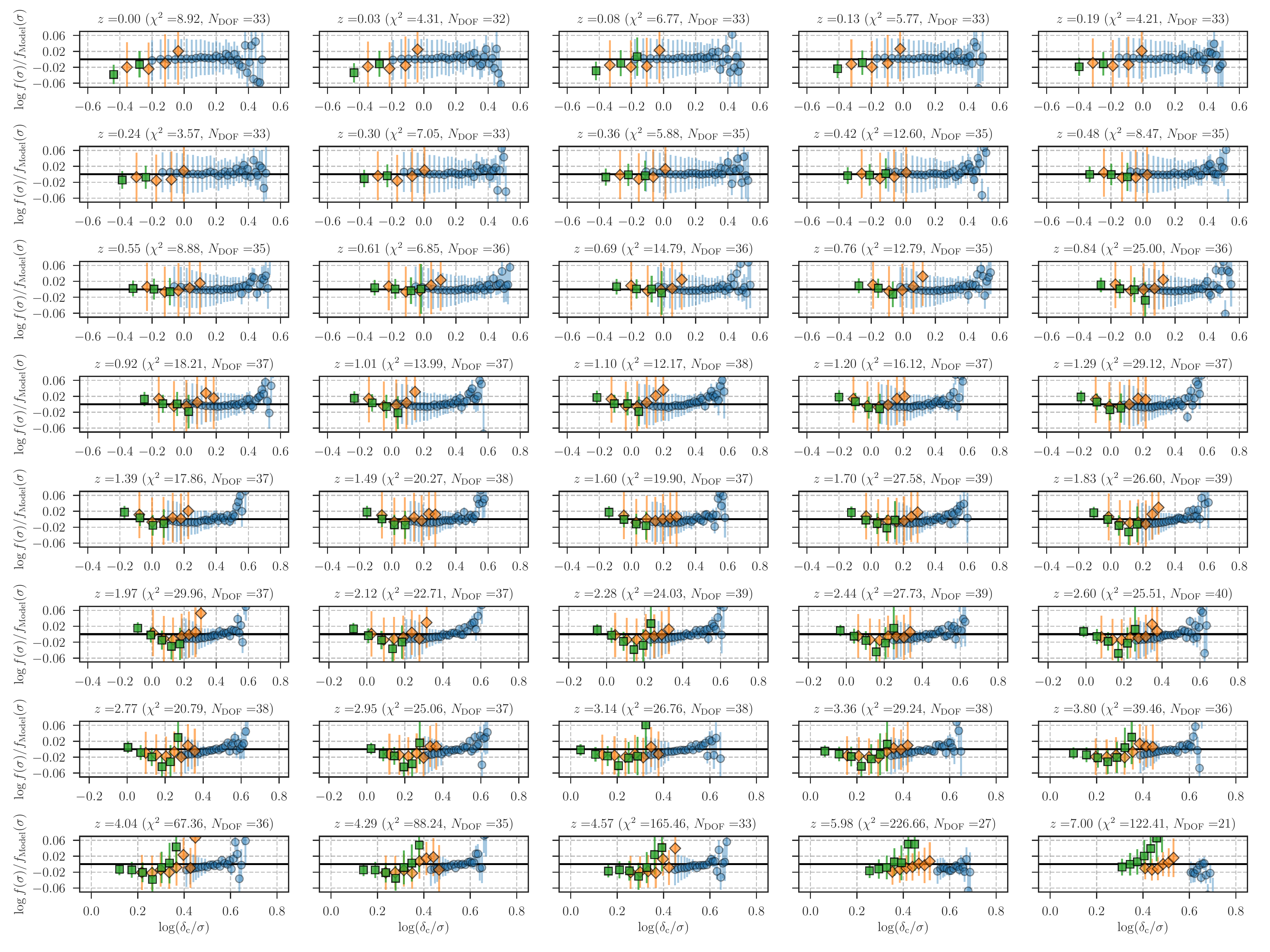}
\caption{
Similar to Figure~\ref{fig:bestfit_fsigma}, but we show the residual between the simulation results and the model with Eq.~(\ref{eq:model_Az})-(\ref{eq:model_qz}).
\label{fig:bestfit_fsigma_model}}
\end{figure*}

In this Appendix, we provide a summary of our fitting results
as in Figure~\ref{fig:bestfit_fsigma}.
Note that Figure~\ref{fig:bestfit_fsigma} shows the residual 
between simulation results and the best-fit model.
It is non-trivial to reproduce the simulation results with a similar level to Figure~\ref{fig:bestfit_fsigma}
when we interpolate the model parameters as in Eq.~(\ref{eq:model_Az})-(\ref{eq:model_qz}).
The comparison with the simulation results and the model with Eq.~(\ref{eq:model_Az})-(\ref{eq:model_qz}) are summarized in Figure~\ref{fig:bestfit_fsigma_model}.
The figure represents the performance of our calibrated model
and we find a 5\%-level precision in our model for a wide range of masses and redshifts.

\bibliographystyle{aasjournal}
\bibliography{sample63}



\end{document}